\documentclass[journal]{IEEEtran}
\usepackage{enumerate}
\usepackage{graphicx}
\usepackage{latexsym}
\usepackage{amsmath}
\usepackage{amssymb}
\usepackage{epsfig}
\usepackage{cite}
\usepackage{algorithmic}
\usepackage{overpic}
\usepackage{multirow}
\usepackage[table]{xcolor}
\usepackage{graphicx}
\usepackage{colortbl,array}
\usepackage{multirow,bigdelim}
\usepackage{graphicx}
\usepackage{subfigure}
\usepackage{booktabs}
\usepackage{booktabs,amsmath}
\usepackage{fancyhdr, graphicx, amsmath, amssymb}
\usepackage[linesnumbered,ruled]{algorithm2e}
\usepackage{arydshln}
\usepackage[english]{babel}
\usepackage{xcolor}
\usepackage{soul}

\usepackage{mathtools}
\newcommand\iidsim{\stackrel{\mathrm{i.i.d.}}{\sim}}

\newtheorem{thm}{Theorem}
\newtheorem{lem}{Lemma}
\newtheorem{rem}{Remark}

\newtheorem{asm}{Assumption}
\newtheorem{defa}{Definition}

\newtheorem{prop}{Proposition}

\begin{document}
%
\title{Finite-Time Model Inference from A Single Noisy Trajectory}

\author{Yanbing~Mao, Naira~Hovakimyan, Petros~Voulgaris, and Lui~Sha
\thanks{Y.~Mao and N.~Hovakimyan are with the Department of Mechanical Science and Engineering, University of Illinois at Urbana--Champaign, Urbana, IL, 61801 USA (e-mail: \{ybmao, nhovakim\}@illinois.edu).}
\thanks{P.~Voulgaris is with the Department of Mechanical Engineering, University of Nevada, Reno, NV, 89557 USA (e-mail: pvoulgaris@unr.edu).}
\thanks{L.~Sha is with the Department of Computer Science, University of Illinois at Urbana--Champaign, Urbana, IL, 61801 USA (e-mail: lrs@illinois.edu).}
\thanks{This work was supported by NSF (award numbers CMMI-1663460, ECCS-1739732 and CNS-1932529).}}

\maketitle

\begin{abstract}
This paper proposes a novel model inference procedure to identify a system from a single noisy trajectory over a finite-time interval.  The proposed inference procedure comprises  an observation data processor, a redundant data processor and an ordinary least-square estimator, wherein the data processors mitigate the influence of observation noise on inference error. We first systematically investigate the comparisons with naive least-square-regression based model inference and uncover that 1) the same observation data has identical influence on the feasibility of the proposed and the naive model inferences, 2) the naive model inference uses all of the redundant data, while the proposed model inference optimally uses the basis and redundant data to mitigate the influence of observation noise on model error. We then study the sample complexity of proposed model inference, wherein observation noise leads to the statistical dependence of samples on time and coordinates. Particularly, we derive the sample-complexity upper bound (on the number of observations sufficient to infer a model with prescribed levels of accuracy and confidence) and the sample-complexity lower bound (high-probability lower bound on model error). Finally, the proposed model inference is numerically validated and analyzed.
\end{abstract}

\begin{IEEEkeywords}
Correlated samples, redundant data, sample complexity, ordinary least-square estimator.
\end{IEEEkeywords}

\IEEEpeerreviewmaketitle

\section{Introduction}
\IEEEPARstart{O}{ne} of the fundamental assumptions of model-based decision making, such as model predictive control \cite{bujarbaruah2018adaptivea,bujarbaruah2018adaptiveb},  stealthy attack and defense strategies of cyber-physical systems \cite{na2018multiplicative} and (mis)-information spreading in social networks \cite{budak2011limiting}, is the availability of a fairly accurate model of the underlying dynamics in consideration. For the systems whose model structure, dimension, parameters and assumptions are sensitive to the operational environment, see e.g., traction control system \cite{borrelli2006mpc} and anti-lock braking system \cite{choi2008antilock} of autonomous vehicles, it is unreasonable to expect that the off-line-built models can accurately capture the dynamics in an unforeseen operational environment, as e.g. the 2019 New York City Snow Squall \cite{software}.
If the unforeseen operational environments or Black Swan events (see e.g., Apollo XIII accident) cause drastic changes of  system dynamics or large deviation from control missions in time- and safety-critical systems, it is critical to timely infer and update the system model for reliable decision making using limited state observation data. These real problems highlight the importance of finite-time model inference or system identification for safe operation in the dynamic, uncertain and unforeseen environments, which aids in fast and flexible response to uncertain and unexpected events. System model inference from observed state data in dynamical systems has been a key problem  in different fields ranging from learning and control \cite{bujarbaruah2018adaptivea,bujarbaruah2018adaptiveb,jain2018learning,nguyen2009local} to bioinformatics \cite{d2000genetic} and social networks \cite{al2011survey}.  We categorize the prior model inference approaches into two groups: approximate and exact model inferences.

\emph{Approximate model inference} from observing the system state is studied in several practical scenarios, where the state observation is only partially available or is stochastic (e.g., noisy), thus reducing the inference to the well-studied problem of estimation of system matrix, with assumptions on the system stability and the distribution of noise, among others. For  distributed and networked dynamical systems, Wiener Filtering \cite{materassi2010topological, materassi2012problem}, structural equation models \cite{ioannidis2019semi} and autoregressive models \cite{nozari2017network} are employed to estimate network topology, leveraging additional tools from estimation theory \cite{nozari2017network}, adaptive feedback control \cite{fazlyab2014robust},  optimization theory with sparsity constraints \cite{hayden2016sparse}, and others. In recent three years, significant effort has been devoted towards the sample-complexity bound of ordinary least-square estimator of system matrix \cite{jedra2020finite,simchowitz2018learning,sarkar2019near,sarkar2019finite,oymak2019non,banerjee2019random}, i.e., the upper bounds on the number of observations sufficient to identify system matrix with prescribed levels of accuracy and confidence (PAC). For example, assuming the process noise vectors are $\mathrm{i.i.d}$ isotropic subgaussian and have $\mathrm{i.i.d}$ coordinates and the real system matrix is stable,  Jedra and Proutiere in \cite{jedra2020finite}
showed that the upper bound matches existing sample complexity lower bounds up to universal multiplicative factors, a result conjectured in \cite{simchowitz2018learning}; Sarkar and Rakhlin in \cite{sarkar2019near} removed the assumption of stable system matrix, and derived the finite-time model error bounds and demonstrated that the ordinary least-squares solution is statistically inconsistent when real system matrix is not regular. We note the sample complexity bounds obtained in \cite{jedra2020finite,simchowitz2018learning,sarkar2019near,sarkar2019finite,oymak2019non} rely on the Hanson-Wright inequality \cite{rudelson2013hanson} that requires zero-mean, unit-variance, sub-gaussian independent coordinates for noise vectors. Banerjee et al. in \cite{banerjee2019random} considered the generalization of existing results by allowing for statistical dependence in stochastic process via the Johnson--Lindenstrauss transform \cite{matouvsek2008variants,ailon2006approximate}, which however still requires the noise variables to have zero mean and the marginal random variables to be conditionally independent.

In this paper, we reveal that even when both process and observation noise in a dynamical system have $\mathrm{i.i.d}$ coordinates and time, the presence of observation noise leads to the dependence of samples on time and coordinates. Moreover, in an adversary environment, the attacker can strategically inject false data to observations for hindering reliable decision making \cite{9099503}, such that the observation noise can have non-zero mean. The statistical dependency and non-zero mean prevent the usage of sample complexity results \cite{jedra2020finite,simchowitz2018learning,sarkar2019near,sarkar2019finite,oymak2019non,banerjee2019random} when observation noise with non-zero/zero mean is injected into dynamical systems. In this paper, we take important strides towards exploring the sample complexity of proposed novel model inference, which can mitigate the influence of observation noise on inference error, through leveraging the variant of the Hanson-Wright inequality \cite{adamczak2015note}.

\emph{Exact model inference} in different settings has been studied as well. For example, Marelli and Fu in \cite{marelli2007exact} proved that with the estimation of input auto-correlation, the exact system model is identified asymptotically as the number of samples approaches infinity, while within the behavioral setting, the
exact system identification is formulated as a Hankel structured low rank matrix completion problem in \cite{markovsky2013exact}. Another required capability for the exact model inference is ``grounding." For example, in \cite{nabi2012network}, with the firstly obtained characteristic polynomial, connecting every two nodes to the ground (set to zero) is required to exactly infer the communication topology of a consensus protocol. In \cite{shahrampour2014topology}, with additional knowledge of eigenvalue-eigenvector of matrix that describes network structure, this ``grounding" approach is coupled with power spectral analysis. We note that while these methods proposed in \cite{marelli2007exact,markovsky2013exact,nabi2012network,shahrampour2014topology} provide exact model inference, it is difficult, if not impossible, to apply them for exact online model inference. This is mainly due  to the requirements of altering real-time state values (``grounding"), infinite sampling, knowledge of partial eigenvalue-eigenvector, etc. A novel exact topology inference procedure is recently developed in\cite{WHT10}, primarily for continuous-time consensus dynamics, by transforming the exact inference problem into a solution of Lyapunov equation whose numerical solutions are well studied, see e.g., \cite{bartels1972solution}. Unlike the ones in \cite{marelli2007exact,markovsky2013exact,nabi2012network,shahrampour2014topology}, this approach does not need the external stimulation, the ``grounding", the infinite sampling, and the knowledge of matrix eigenvalue-eigenvector. This approach works mainly for continuous-time consensus dynamics with undirected communication. In other words, it is not sufficient for exact inference for the directed network topology, or dynamical system with asymmetric system matrix. Additionally, this approach needs the knowledge of system initial conditions. These traits and requirements hinder its application to finite-time online model inference for more general dynamical systems.

In this paper, both the proposed and defined naive model inference procedures can exactly infer system matrix when observation and process noises are time-invariant, i.e., have zero variances. Unlike the one in \cite{WHT10}, the two approaches do not need the knowledge of initial condition. More importantly, they can exactly infer the asymmetric system matrix and the processed bias in observed trajectory.

Our contributions are summarized as follows:
\begin{itemize}
  \item We propose a novel model inference procedure that comprises  an observation data processor, a redundant data processor and an ordinary least-square estimator, which can significantly mitigate the influence of observation noise on the model error.
  \item We investigate the systematic comparisons between the proposed and defined naive model inference procedures and discover that
  \begin{itemize}
    \item the same observation data has identical influence on the inference feasibility of the proposed and naive model inferences;
    \item the naive model inference uses all of the redundant data, while the proposed model inference optimally uses redundant and basis data to reduce the model error.
  \end{itemize}
    \item We derive the sample-complexity upper bound on the number of observations for PAC and the high-probability lower bound on the model error, which allows the observation noise to have non-zero mean and induce the statistical dependence of samples on coordinates and observation time.
    \item Leveraging the derived sample-complexity upper bound, we provide an algorithm of PAC verification and updating. The upper bound is leveraged by the redundant data processor of inference procedure to reduce model error.
\end{itemize}

This paper is organized as follows. In Section II, we present the preliminaries. Section III presents the proposed and naive model inference procedures, as well as their comparisons. In Section VI, we investigate sample complexity. The sample-complexity bounds and PAC verification are studied in Section V. We next present our numerical results in Section VI. Section VII presents our conclusions and future research directions.

\vspace{-0.00cm}
\section{Preliminaries}
\vspace{-0.1cm}
\subsection{Notation}
 We use $P$ $\leq$ 0 to denote a negative semi-definite matrix $P$. We let $\mathbb{R}^{n}$ and $\mathbb{R}^{m \times n}$ denote the set of $\emph{n}$-dimensional real vectors and the set of $m \times n$-dimensional real matrices, respectively. $\mathbb{N}$ stands for the set of natural numbers, and $\mathbb{N}_{0} = \mathbb{N} \cup \{0\}$. The superscript `$\top$' stands for the matrix transposition. $\left[\mathbf{x}\right]_{i}$ denotes the $i$th entry of a vector $\mathbf{x}$.
 We let $\mathbf{1}_{m}$ and $\mathbf{0}_{m}$, respectively, denote the $m$-dimensional vectors of all ones and all zeros.
We define $\mathbf{I}_{n}$ as $n \times n$-dimensional identity matrix. We define $\mathbf{O}_{m \times n}$ as $m \times n$-dimensional zero matrix. The symmetric terms in a matrix are denoted by $\ast$. $|| \cdot ||$ denotes  the spectral norm of a matrix, or the Euclidean norm of a vector. Other notations are highlighted as follows:
\begin{description}
  \item[$|\mathbb{T}|:$] ~~ cardinality (i.e., size) of set $\mathbb{T}$;
  \item[$|| A ||_{\mathrm{F}}:$] ~~ Frobenius norm of matrix $A$;
  \item[$\text{tr}(A):$] ~~ trace of matrix $A$;
  \item[$\mathbf{E}$] ~~  expectation operator;
  \item[$\mathcal{S}^{n-1}:$] ~~  unit sphere in $\mathbb{R}^{n}$;
  \item[$\Omega^{\mathrm{c}}:$] ~~~complement of event $\Omega$;
  \item[$\mathbf{P}(\Omega):$] ~~~probability of event $\Omega$.
\end{description}

\vspace{-0.1cm}
\subsection{Problem Formulation}
In this paper, we use the following dynamics to describe the behavior of real plants in unknown operational environment:
\begin{subequations}
\begin{align}
\mathbf{x}(k+1) &= A\mathbf{x}(k) + \mathbf{a} + \mathbf{f}(k), \label{realdynhh}\\
\mathbf{r}(k)  &= \mathbf{x}(k) + \mathbf{w}(k), ~~~~k \in \mathbb{N}
\end{align}\label{realdyna}
\end{subequations}
\!\!where $\mathbf{x}(k) \in \mathbb{R}^{n}$ is the system state, $\mathbf{a} \in \mathbb{R}^{n}$, $A \in \mathbb{R}^{n \times n}$ is the system matrix, $\mathbf{r}(k)$ is the state observation, $\mathbf{w}(k)$ denotes the observation noise vector, $\mathbf{f}(k)$ denotes the process noise that can represent the nonlinear factor, model approximation error and uncertainty.

For model inference/learning, the following two metrics are defined to assess the quality of inferred/learned model:
\begin{itemize}
  \item Generalization Error: it measures how well the model fits unseen data.
  \item Model Error: it measures how far the identified model is from the ``true" model.
\end{itemize}
Due to the observation noise, small generalization error does not necessarily mean small model error, and vice versa. Specifically, with strategically injected observation noise, a well identified model with smaller model error can have larger generalization error, i.e., it poorly fits the unseen (corrupted) observation data. For this reason, we use model error to assess the quality of inferred model, with a focus on
\begin{description}
  \item[O1:] mitigating the influence of observation noise on model error via incorporating data processors into model inference procedure;
  \item[O2:] sample-complexity bounds for the proposed model inference solution to be $(\phi, \delta)$-PAC: prescribed levels $(\phi, \delta)$ of accuracy and confidence, i.e., $\mathbf{P} (e_{M} \leq \phi) \geq 1 - \delta$, where $\phi > 0$, $0 < \delta < 1$ and $e_{M}$ denotes model inference error.
\end{description}

Recently, sample complexity of an ordinary least-square estimator in identifying system matrix $A$ (assuming $\mathbf{a} = \mathbf{0}_{n}$) have been studied in \cite{jedra2020finite,simchowitz2018learning,sarkar2019near,sarkar2019finite,oymak2019non,banerjee2019random}. To review whether the obtained bounds therein can be applied to the dynamical systems \eqref{realdyna}, let us consider the dynamics of observation obtained from \eqref{realdyna}:
\begin{subequations}
\begin{align}
\mathbf{r}(k+1) &= A\mathbf{r}( k) + \mathbf{h}(k+1), ~~~\mathbf{r}(1) = \mathbf{x}(1) + \mathbf{w}(1) \label{undb}\\
\mathbf{h}(k+1) &= \mathbf{a} + \mathbf{f}(k) + \mathbf{w}(k+1) - A\mathbf{w}(k), ~~k \in \mathbb{N}. \label{undbred}
\end{align}\label{un}
\end{subequations}
\!\!To distinguish from the observation noise $\mathbf{w}(k)$ and process noise $\mathbf{f}(k)$ in the real system \eqref{realdyna}, we refer to $\mathbf{h}(k)$ in the dynamics \eqref{un} as \emph{processed bias}.

\begin{rem}[Statistical Dependencies] Assuming both $\mathbf{f}(k)$ and $\mathbf{w}(k)$ are random vectors, as indicated by \eqref{undbred}, the term $\mathbf{w}(k+1) - A\mathbf{w}(k)$ induces the dependence of ${\left\{ {\mathbf{h}\left( k \right)} \right\}_{k \in \mathbb{N}}}$ and conditional dependence of ${\left\{ {\left. {\mathbf{r}\left( k \right)} \right|\mathbf{w}\left( k \right),\mathbf{f}\left( k \right)} \right\}_{k \in \mathbb{N}}}$ on coordinates and observation time. We note that the sample complexity analysis in \cite{jedra2020finite,simchowitz2018learning,sarkar2019near,sarkar2019finite,oymak2019non} relies on the assumption that ${\left\{ {\mathbf{h}\left( k \right)} \right\}_{k \in \mathbb{N}}}$ is $\mathrm{i.i.d}$ and the coordinates of $\mathbf{h}(k)$ are $\mathrm{i.i.d}$, while the analysis in \cite{banerjee2019random} assumes that ${\left\{ {r\left( k \right)} \right\}_{k \in \mathbb{N}}}$ has conditional independence. Therefore, the derived sample-complexity bounds in \cite{jedra2020finite,simchowitz2018learning,sarkar2019near,sarkar2019finite,banerjee2019random} cannot be applied to the dynamical system \eqref{un} or \eqref{realdyna}.
\end{rem}

To achieve the objectives O1 and O2, we first process the observation data in the following way, which corresponds to the Observation Data Processor in the model inference procedure, as shown in Figure \ref{modelinfer} (a):
\begin{align}
\mathbf{r}^{q}_{m} &\triangleq \mathbf{r}( m ) - \mathbf{r}( q ), ~~m < q \in \mathbb{N}, \label{deff1}
\end{align}
whose dynamics are obtained from \eqref{un} as
\begin{align}
&{\bf{r}}^{q+1}_{m+1} \!=\! A{\bf{r}}^{q}_{m} \!+\! {\bf{h}}^{q+1}_{m\!+\!1}, ~\mathbf{h}^{q+1}_{m+1} \!\triangleq\! \mathbf{h}( m\!+\!1 ) \!-\! \mathbf{h}(q+1 ).  \label{pam0}
\end{align}

\begin{rem}[Trajectories] We obtain the trajectory of observations from the real system \eqref{realdyna} as
\begin{subequations}
\begin{align}
\!\mathbf{r}( k ) &= A^{k-1}\mathbf{x}(1) +  \mathbf{w}( k ) + \sum\limits_{i = 0}^{k - 2} {{A^i}} (\mathbf{f}(k \!-\! 1 \!-\! i)+\mathbf{a}),\\
\!\mathbf{r}( 1 ) &= \mathbf{w}( 1 ) + \mathbf{x}( 1 ),
\end{align}\label{unda}
\end{subequations}
by which (and using \eqref{deff1}), we arrive at
\begin{align}
&\mathbf{r}_k^m = \left( {{A^{k - 1}} \!-\! {A^{m - 1}}} \right)\!\mathbf{x}(1) + \mathbf{w}_k^m + \sum\limits_{i = 0}^{k - 2} \!{A^{i}}{\mathbf{f}_{k - 1- i}^{m - 1 -  i}} \nonumber \\
&\hspace{2.30cm} - \sum\limits_{i = k - 1}^{m - 2} \!\!\!{{A^i}(\mathbf{f}(m \!-\! 1 \!-\! i)}\!+\!\mathbf{a}), \!~m \!>\! k \!\in\! \mathbb{N} \label{dv1}
\end{align}
where $\mathbf{f}^{m}_{k} \triangleq \mathbf{f}( k ) - \mathbf{f}( m )$ and $\mathbf{w}^{m}_{k} \triangleq \mathbf{w}( k ) - \mathbf{w}( m )$.
\end{rem}

\begin{rem}[Observation Data Processor]
If the observation noise has zero variance, i.e., $\mathbf{w}(k) \equiv \mathbf{w}$, it follows from \eqref{undbred} and  \eqref{deff1} that $\mathbf{h}^{q+1}_{m+1} = \mathbf{f}(m+1) - \mathbf{f}(q+1)$, such that the bias $\mathbf{h}^{q+1}_{m+1}$ in system \eqref{pam0} is relevant to the process noise $\mathbf{f}(k)$ only, which is \emph{one} motivation behind the data processor \eqref{deff1} in mitigating the influence of the observation noise with the non-zero mean on the model error.
\end{rem}

For the observations used for inference in a time interval $\{k, k + 1, \ldots, p-1,p\}$, we define two \emph{increasingly ordered} (possibly, empty) sets $\mathbb{T}_{k}$ and $ \mathbb{L}_{k}$, such that
\begin{align}
\mathbb{T}_{k} \subseteq \mathbb{L}_{k} \triangleq \left\{ {k+1,~k+2,~\ldots,~p-1,~p} \right\}.\label{deffkh1}
\end{align}
We note that the relation \eqref{deffkh1} indicates that $\mathbb{T}_{k+r}$ $\subseteq$ $\mathbb{L}_{k+r} \!=\! \left\{ {k\!+\!r\!+\!1,k\!+\!r\!+\!2,\ldots,p\!-\!1,p} \right\}$, for $\forall r \in \mathbb{N}_{0}$. Based on \eqref{deffkh1}, we construct the following data matrices:
\begin{subequations}
\begin{align}
&\breve{X}_{k} \triangleq  \begin{cases}
\left[ {\mathbf{r}^{\mathbb{T}_k(1)}_{k},~\mathbf{r}^{\mathbb{T}_k(2)}_{k}, ~\ldots, ~\mathbf{r}^{\mathbb{T}_k(\left| {{\mathbb{T}_k}} \right|)}_{k}} \right], &\text{if}~\mathbb{T}_k \neq \varnothing\\
`\mathrm{null}',&\text{otherwise}
\end{cases}\label{sufk1}\\
&\breve{X}_{(k,p)} \triangleq  \left[ {\breve{X}_{k}, ~\breve{X}_{k+1}, ~\ldots, ~\breve{X}_{p-1}} \right],\label{sufk2}
\end{align}\label{ddvs}
\end{subequations}
\!\!\!where $\mathbb{T}_k(m)$ denotes the $m$th element of (non-empty) \emph{increasingly ordered} set $\mathbb{T}_k$, and thus, $m \in \{1,\ldots,\left| {{\mathbb{T}_k}} \right|-1,\left| {{\mathbb{T}_k}} \right|\}$.

\section{Model Inference Procedures}
\begin{figure}[http]
\centering{
\includegraphics[scale=0.60]{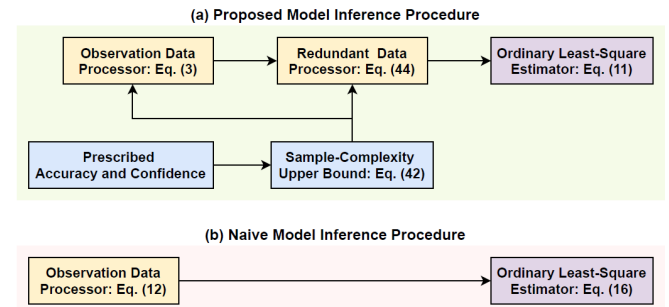}
}
\caption{Proposed and naive model inference architectures: naive one takes all of the redundant data into account.}
\label{modelinfer}
\end{figure}

In this section, we present two model inference procedures: \emph{Proposed Model Inference} and \emph{Naive Model Inference} (whose architectures are shown in Figure \ref{modelinfer} (a) and (b), respectively), and their systematic comparisons.

\subsection{Proposed Model Inference}
\vspace{-0.1cm}
We now present the critical relation of data matrices in the following lemma.
\begin{lem}
Consider the system \eqref{pam0} and matrix \eqref{ddvs}. We have
\begin{align}
AP_{(k,p)} = Q_{(k,p)} - R_{(k,p)}, \label{lm3r}
\end{align}
where we define:
\begin{subequations}
\begin{align}
{P_{\left( {k,p} \right)}} &\!\triangleq\! \sum\limits_{m = k}^{p - 1} {\sum\limits_{q = \mathbb{T}_m(1)}^{\mathbb{T}_m(\left| {{\mathbb{T}_m}} \right|)} {\mathbf{r}_m^q{{\left( {\mathbf{r}_m^q} \right)^\top}}}} = \breve{X}_{(k,p)}\breve{X}^{\top}_{(k,p)}, \label{defmatbdis}\\
{Q_{\left( {k,p} \right)}} &\!\triangleq\! \sum\limits_{m = k}^{p - 1} {\sum\limits_{q = \mathbb{T}_m(1)}^{\mathbb{T}_m(\left| {{\mathbb{T}_m}} \right|)} \!\!{\mathbf{r}_{m + 1}^{q+1}{{\left( {\mathbf{r}_m^q} \right)^\top}}} } \!\!=\! \breve{X}_{(k+1,p+1)}\breve{X}^{\top}_{(k,p)}, \label{defmatbdis1}\\
{R_{\left( {k,p} \right)}} &\!\triangleq\! \sum\limits_{m = k}^{p - 1} {\sum\limits_{q = \mathbb{T}_m(1)}^{\mathbb{T}_m(\left| {{\mathbb{T}_m}} \right|)} {\mathbf{h}_{m+1}^{q+1}{{\left( {\mathbf{r}_m^q} \right)^\top}}} }. \label{defmatbdis2}
\end{align}
\end{subequations}
\end{lem}

\begin{IEEEproof}
Pre-multiplying the both sides of \eqref{defmatbdis} by $A$ yields
\begin{align}
A{P_{\left( {k,p} \right)}} &\triangleq \sum\limits_{m = k}^{p - 1} {\sum\limits_{q = \mathbb{T}_m(1)}^{\mathbb{T}_m(\left| {{\mathbb{T}_m}} \right|)} A{\mathbf{r}_m^q{{\left( {\mathbf{r}_m^q} \right)^\top}}} }, \nonumber
\end{align}
substituting \eqref{pam0} into which directly results in
\begin{align}
A{P_{\left( {k,p} \right)}} &\triangleq \sum\limits_{m = k}^{p - 1} {\sum\limits_{q = \mathbb{T}_m(1)}^{\mathbb{T}_m(\left| {{\mathbb{T}_m}} \right|)} {({\bf{r}}^{q+1}_{m+1} - {\bf{h}}^{q+1}_{m+1}){{\left( {\mathbf{r}_m^{q}} \right)^\top}}} }, \nonumber
\end{align}
which, in conjunction with \eqref{defmatbdis1} and \eqref{defmatbdis2}, yields \eqref{lm3r}.
\end{IEEEproof}

We note that due to unknown ${\bf{h}}^{{q+1}}_{m+1}$, the matrix $R_{(k,p)}$ in relation \eqref{lm3r} is also unknown. We obtain the following model inference solution which corresponds to the Ordinary Least-Square Estimator in the proposed model inference procedure as shown in Figure \ref{modelinfer} (a):
\begin{subequations}
\begin{align}
A_{\text{if}} &= Q_{(k,p)}{P^{ - 1}_{(k,p)}},~\text{provided}~\text{rank}(P_{(k,p)}) = n, \label{self}\\
\mathbf{a}_{\text{if}} &= \frac{1}{{p - k}}\sum\limits_{m = k}^{p-1} {( {\mathbf{r}(m + 1) - {A_{\text{if}}}\mathbf{r}(m)})}.\label{selfaa}
\end{align}\label{selffn}
\end{subequations}

\vspace{-0.4cm}
\subsection{Naive Model Inference}
\vspace{-0.1cm}
We now consider an alternative inference procedure: naive model inference, as shown in Figure \ref{modelinfer} (b), whose Observation Data Processor corresponding to \eqref{undb} is
\begin{subequations}
\begin{align}
\!\!\widehat{Y}_{(k,p)} &\triangleq \left[{\bf{r}}( {k \!+\! 1}),~{\bf{r}}( {k \!+\! 2}), ~\ldots, ~{\bf{r}}(p)\right]^{\top},\label{xf1}\\
\!\!\widehat{X}_{(k,p)} &\triangleq \left[ {\left[ \begin{array}{l}
\!\!{\bf{r}}(k)\!\!\!\!\\
\!\!1\!\!\!\!
\end{array} \right]\!,~\left[ \begin{array}{l}
\!\!{\bf{r}}(k + 1)\!\!\!\!\\
\!\!1\!\!\!\!
\end{array} \right]\!, ~\ldots,~\left[ \begin{array}{l}
\!\!{\bf{r}}(p-1)\!\!\!\!\\
\!\!\!1\!\!\!\!
\end{array} \right]} \right]^{\top}\!.\label{xf2}
\end{align}\label{zgg}
\end{subequations}
Correspondingly, we define:
\begin{subequations}
\begin{align}
\!\!\widehat{A} &\triangleq \left[ A, ~\mathbf{a}\right], \label{xf3}\\
\!\!\widehat{E}_{(k,p)} &\triangleq \left[{\bf{h}}(k\!+\!1) \!-\! \mathbf{a}, ~{\bf{h}}(k\!+\!2) \!-\! \mathbf{a}, ~\ldots, ~{\bf{h}}(p) \!-\! \mathbf{a}\right],\label{xf}
\end{align}\label{zgg}
\end{subequations}
With the definitions at hand, we obtain from system \eqref{un} that
\begin{align}
\widehat{Y}^{\top}_{(k,p)} = \widehat{A}\widehat{X}^{\top}_{(k,p)} + \widehat{E}_{(k,p)}.   \label{datanew}
\end{align}

The system identification problem is reduced to estimating $\widehat{A}$ based on the available data matrices $\widehat{Y}_{(k,p)}$ and $\widehat{X}_{(k,p)}$. We then consider the Ordinary Least-Square Estimator:
\begin{align}
{\widehat{A}^\top_{\emph{\emph{ls}}}} = \mathop {\arg \min }\limits_{\widehat{A}^\top \in  \mathbb{R}^{(n+1) \times n}} || {\widehat{Y}_{(k,p)} - \widehat{X}_{(k,p)}{\widehat{A}^\top}} ||_2^2, \label{lestobj}
\end{align}
whose optimal solution has been obtained in \cite{kay1993fundamentals} as
\begin{align}
&\widehat{A}_{\emph{\emph{ls}}}^\top = {(\widehat{X}^{\top}_{(k,p)}\widehat{X}_{(k,p)})^{ - 1}}\widehat{X}^{\top}_{(k,p)}\widehat{Y}_{(k,p)}, \nonumber\\
&\hspace{2.4cm}\text{provided}~\text{rank}(\!\widehat{X}^{\top}_{(k,p)}\widehat{X}_{(k,p)}) = n+1. \label{les}
\end{align}

\vspace{-0.5cm}
\subsection{Inference Feasibility and Solution}
We formally define the feasibility of the proposed and naive model inferences as follows.
\begin{defa}
The proposed model inference solution \eqref{selffn} is said to be feasible if and only if $\text{rank}\left(P_{(k,p)}\right) = n$, while the naive model inference solution \eqref{les} is said to be feasible if and only if $\text{rank}(\widehat{X}^{\top}_{(k,p)}\widehat{X}_{(k,p)}) = n + 1$. \label{dfaa}
\end{defa}

We then present auxiliary results in the following lemmas pertaining to inference feasibility.
\begin{lem}
Consider the data matrices \eqref{sufk1}, \eqref{defmatbdis} and \eqref{xf2}, and the set \eqref{sufk1}.  We have
\begin{subequations}
\begin{align}
&\ker(P_{(k,p)}) = \ker(\breve{X}^{\top}_{k}), ~~~~~~\text{if}~\left| {{\mathbb{T}_k}} \right| = p-k\label{reslem2a}\\
&\ker(\widehat{X}^{\top}_{(k,p)}\widehat{X}_{(k,p)}) = \ker(\widehat{X}_{(k,p)}). \label{reslem2b}
\end{align}
\end{subequations}\label{ko}
\end{lem}
\vspace{-0.5cm}
\begin{IEEEproof}
See Appendix B.
\end{IEEEproof}

\begin{lem}
If the proposed model inference solution \eqref{selffn} is not feasible, the naive model inference solution \eqref{les} is not feasible as well.
\label{cr1}
\end{lem}

\begin{IEEEproof}
See Appendix C.
\end{IEEEproof}

In light of Lemmas \ref{ko} and \ref{cr1}, we obtained the comparisons of the two inferences, which are formally presented in the following theorem.
\begin{thm}
Consider the proposed model inference solution \eqref{selffn} with data matrix \eqref{defmatbdis}, and the naive model inference solution \eqref{les}.
\begin{enumerate}
  \item The same observation data has identical influence on the feasibility of the proposed inference solution \eqref{selffn} and the naive model inference solution \eqref{les}, i.e,
  \begin{itemize}
  \item $\text{rank}(P_{(k,p)}) \!=\! n$, if and only if $\text{rank}(\widehat{X}^{\top}_{(k,p)}\widehat{X}_{(k,p)}) \!=\! n \!+\! 1$;
  \item $\text{rank}(P_{(k,p)}) \!<\! n$, if and only if $\text{rank}(\widehat{X}^{\top}_{(k,p)}\widehat{X}_{(k,p)}) \!<\! n \!+\! 1$.
\end{itemize}
  \item The two inference procedures generate the same inference solution, if
\begin{align}
\!\text{rank}({{P}_{(k,p)}}) \!=\! n, ~~{P}_{(k,p)} \!=\! \sum\limits_{m = k}^{p - 1}{\sum\limits_{m < q = k + 1}^{p}\!\!\!{\mathbf{r}^{q}_{m}{{(\mathbf{r}^{q}_{m})^\top}}} }\!. \label{npp}
\end{align}
\end{enumerate}\label{fthmp}
\end{thm}
\begin{IEEEproof}
See Appendix D.
\end{IEEEproof}

\begin{rem} [Exact Inference of System Matrix]
If the process noise $\mathbf{f}(k)$ and the observation noise $\mathbf{w}(k)$ have zero variance, i.e., $\mathbf{f}(k) \equiv \mathbf{f}$ and $\mathbf{w}(k) \equiv \mathbf{w}$, it follows from \eqref{undbred} and the definition of $\mathbf{h}^{q+1}_{m+1}$ in \eqref{pam0} that $\mathbf{h}^{q+1}_{m+1} \equiv \mathbf{0}_{n}$, such that ${R_{\left( {k,p} \right)}} = \mathbf{O}_{n \times n}$. We then conclude from \eqref{lm3r} and the statement 2) of Theorem \ref{fthmp} that both the proposed inference \eqref{selffn} and the naive model inference \eqref{les} can exactly system matrix $A$, i.e., $A_{\text{if}} = A  = A_{\text{ls}}$.
\end{rem}

To explore the implicit insight of Theorem \ref{fthmp}, we introduce a definition to classify the processed observation data.
\begin{defa}  [Basis and Redundant Data]
Given the observation data $\mathbf{r}(m)$ in a time interval $\{k, k + 1, \ldots, p\}$, any processed data in a set $\mathbb{X}_k^p \triangleq \left\{ {\left. {\mathbf{r}_m^q} \right|k \le m < q \leq p} \right\}$ is called \emph{basis data}, if all of the data vectors in $\mathbb{X}_k^p $ are linearly independent and $\left| {\mathbb{X}_k^p} \right| = p - k$,
while all other data vectors not included in $\mathbb{X}_k^p$ are called to be \emph{redundant data}. \label{def22}
\end{defa}

\begin{rem}
The two statements in Theorem \ref{fthmp}, respectively, indicate that
\begin{itemize}
  \item redundant data has no influence on inference feasibility, while it can influence only model error, consequently, $(\phi, \delta)$-PAC, which motivates the Redundant Data Processor as shown in Figure \ref{modelinfer} (a);
  \item naive model inference takes all of the redundant data into estimator, which explains the lack of Redundant Data Processor in the naive model inference procedure shown in Figure \ref{modelinfer} (b).
\end{itemize}
As shown in Figure \ref{modelinfer} (a), the data processors of the proposed model inference need the knowledge of the sample-complexity upper bound on the number of observations that is sufficient for the inference solution to be $(\phi, \delta)$-PAC. In the following sections, we study the sample complexity of proposed model inference with its associated bounds. \label{kkc}
\end{rem}

\vspace{-0.0cm}
\section{Sample Complexity}
\vspace{-0.1cm}
As indicated by Figure \ref{modelinfer} (a), we need to investigate its sample complexity to complete the whole inference architecture. In this section, we first present its assumptions, then the notations obtained under the assumptions, and finally the sample complexity.
\vspace{-0.5cm}
\subsection{Assumptions}
\vspace{-0.1cm}
Based on the dynamics \eqref{dv1}, we define the following stacked vector, which will be used to describe the Assumptions.
\begin{align}
\eta_{(k,p)} \triangleq [\breve{\mathbf{x}}^{\top}(1), ~~{{\widehat{\bf{w}}}^{\top}_{( {k,p})}}, ~~\breve{\mathbf{f}}_{(k,p)}^\top,~~  \widetilde{\mathbf{f}}_{(k,p)}^\top]^{\top},\label{ghbvector}
\end{align}
where
\vspace{-0.1cm}
\begin{subequations}
\begin{align}
&\breve{\mathbf{x}}(1) \triangleq \left[ {{\mathbf{x}^\top}(1),~{\mathbf{x}^\top}(1), ~\ldots, ~{\mathbf{x}^\top}(1)} \right]\!^{\top} \in {\mathbb{R}^{\sum\limits_{m = k}^{p - 1} \!\!{\left| {{\mathbb{T}_m}} \right|n} }},\label{ghb1}\\
&{\widehat{\mathbf{w}}}_{k} \triangleq  \begin{cases}
[ {{{( {{\bf{w}}_k^{\mathbb{T}_k(1)}})^\top}}\!, \ldots, {{( {{\bf{w}}_k^{\mathbb{T}_k(\left| {{\mathbb{T}_k}} \right|)}})^\top}}} ]\!, \!\!\!&\text{if}~\mathbb{T}_k \!\neq\! \varnothing\\
`\mathrm{null}',\!\!\!&\text{otherwise}
\end{cases}\label{ghb2}\\
&{{\widehat{\bf{w}}}_{( {k,p})}} \triangleq [ {{\widehat{\bf{w}}}_k,~ {\widehat{\bf{w}}}_{k + 1},~ \ldots,~ {\widehat{\bf{w}}}_{p - 2},~ {\widehat{\bf{w}}}_{p - 1}}]^{\top},\label{ghb3}\\
&\breve{\mathbf{f}}^{m}_{k} \triangleq  [ {{{( {\mathbf{f}_{k - 1}^{m - 1}})^\top}}, ~{{( {\mathbf{f}_{k - 2}^{m - 2}})^\top}},~ \ldots, ~{{( {\mathbf{f}_1^{m - k + 1}})^\top}}}],\label{ghb4}\\
&\breve{\mathbf{f}}_k \triangleq  \begin{cases}
[ {{{{\breve{\mathbf{f}}_{k}^{\mathbb{T}_k(1)}}}},~{{{\breve{\mathbf{f}}_{k}^{\mathbb{T}_k(2)}}}}, ~\ldots, ~{{{\breve{\mathbf{f}}_k^{\mathbb{T}_k(\left| {{\mathbb{T}_k}} \right|)}}}}}], ~~&\text{if}~\mathbb{T}_k \!\neq\! \varnothing\\
`\mathrm{null}',~~&\text{otherwise}
\end{cases}\label{ghb4}\\
&\breve{\mathbf{f}}_{(k,p)} \triangleq [ {\breve{\mathbf{f}}_k,~\breve{\mathbf{f}}_{k + 1}, ~\ldots ,~\breve{\mathbf{f}}_{p - 2},~\breve{\mathbf{f}}_{p - 1}}]^{\top},\label{ghb5}\\
&\widetilde{\mathbf{f}}_k^m \triangleq [ {{\mathbf{f}^\top}\!(m \!-\! k)\!+\!\mathbf{a}^\top, \!~\ldots, \!~{\mathbf{f}^\top}(2)\!+\!\mathbf{a}^\top,\!~{\mathbf{f}^\top}(1)\!+\!\mathbf{a}^\top}],\label{ghb6}\\
&\widetilde{\mathbf{f}}_k \triangleq  \begin{cases}
[ {\widetilde{\mathbf{f}}_k^{\mathbb{T}_k(1)},~\widetilde{\mathbf{f}}_k^{\mathbb{T}_k(2)}, ~\ldots, ~\widetilde{\mathbf{f}}_k^{\mathbb{T}_k(\left| {{\mathbb{T}_k}} \right|)}} ], ~&\text{if}~\mathbb{T}_k \!\neq\! \varnothing\\
`\mathrm{null}',~&\text{otherwise}
\end{cases}\label{ghb6}\\
&\widetilde{\mathbf{f}}_{(k,p)} \triangleq [ {\widetilde{\mathbf{f}}_k,~~\widetilde{\mathbf{f}}_{k + 1},~~ \ldots ,~~\widetilde{\mathbf{f}}_{p - 2},~~\widetilde{\mathbf{f}}_{p - 1}}]^{\top},\label{ghb7}
\end{align}\label{ghb}
\end{subequations}
\vspace{-0.3cm}

We introduce the following definition, which will be used to describe the assumptions made on noise.
\begin{defa}[Convex Concentration Property \cite{adamczak2015note}] Let $\mathbf{f}$ be a random vector in $\mathbb{R}^{n}$. We will say that $\mathbf{f}$ has the convex concentration property with constant $\kappa$, if for every
1-Lipschitz convex function $\varphi: \mathbb{R}^{n} \rightarrow \mathbb{R}$, we have $\mathbf{E}|\varphi(\mathbf{f})| < \infty$, and for every $t > 0$ the following holds:
\vspace{-0.1cm}
\begin{align}
\mathbf{P}\left( {\left| {\varphi( \mathbf{f} )} - \mathbf{E}[{\varphi ( \mathbf{f} )}] \right| \ge t} \right) < 2{e^{ - \frac{{{t^2}}}{{{\kappa^2}}}}}. \nonumber
\end{align}
\end{defa}

We now make the following assumption on the initial condition, observation and process noise throughout this paper.
\begin{asm}
Consider the real system \eqref{un} with processed bias $\mathbf{h}^{m}_{k}$ given in \eqref{pam0} and $\eta_{( {k,p})}$ given by \eqref{ghbvector}.
\begin{enumerate}
  \item $\mathbf{f}(k)$ $\iidsim$ $\mathcal{D}_{\mathrm{p}}(\mathbf{0}_{n},\sigma^{2}_{\mathrm{p}}\mathbf{I}_{n})$, ~~$\mathbf{w}(k)$ $\iidsim$ $\mathcal{D}_{\mathrm{o}}(\mu\mathbf{1}_{n},\sigma^{2}_{\mathrm{o}}\mathbf{I}_{n})$, ~~~~~~~~~~~$\mathbf{x}(1)$ $\sim$ $\mathcal{D}_{\mathrm{i}}(\mathbf{0}_{n},\sigma^{2}_{\mathrm{i}}\mathbf{I}_{n})$, ~~~~~~~~$\mathbf{a}$ $\sim$ $\mathcal{D}_{\mathrm{a}}(\mathbf{0}_{n},\sigma^{2}_{\mathrm{a}}\mathbf{I}_{n})$;
  \item $\eta_{( {k,p})}$ has the convex concentration property with constant $\kappa > 0$;
  \item $[\mathbf{h}^{m}_{k}]_{i}$, $i = 1, \ldots, n$,  is $\mathcal{F}_{k}$-measurable and conditionally $\gamma$-sub-Gaussian for some $\gamma$ $>$ $0$, i.e., $\mathbf{E}[ {{e^{\left. {\lambda {[\mathbf{h}^{m}_{k+1}]_{i}}} \right|{\mathcal{F}_k}}}} ] \le {e^{{\frac{\lambda^2\gamma^2}{2}}}}$ for all $\lambda \in \mathbb{R}$.
\end{enumerate}\label{asprv}
\end{asm}

\begin{rem}
Examples of $\eta_{( {k,l})}$ under Assumption \ref{asprv}-2) include any random vector $\eta \in \mathbb{R}^{g}$ with independent coordinates and almost sure $|[\eta]_{i}| \leq 1$ for any $i \in \{1, \ldots, g\}$ \cite{samson2000concentration}, random vectors obtained via sampling without replacement \cite{adamczak2016circular}, vectors with bounded coordinates satisfying some uniform mixing conditions or Dobrushin type criteria \cite{samson2000concentration}, among others \cite{adamczak2015note}. Under Assumption \ref{asprv}-3),  examples of $[\mathbf{h}^{m}_{k}]_{i}$ include a bounded zero-mean noise lying in an interval of length at most $2\gamma$ \cite{abbasi2011improved}, a zero-mean Gaussian noise with variance at most $\gamma^{2}$ \cite{abbasi2011improved}, among many others.
\end{rem}

Based on Assumption, we next present some obtained notations for the presentation of sample complexity.
\vspace{-0.3cm}
\subsection{Obtained Notations}
Under Assumption \ref{asprv}-1), we obtain from \eqref{dv1} that
\begin{align}
&{\bf{E}}\left[\mathbf{r}_k^m(\mathbf{r}_k^m)^{\top}\right] = {\Phi_{(k,m)}} - {\Omega_{(k,m)}} - \Theta_{(k,m)},\label{kop3zd}
\end{align}
where we define:
\begin{subequations}
\begin{align}
&{\Omega_{(k,m)}}  \\
&\!\triangleq\! \left\{ \begin{array}{l}
\!\!\sum\limits_{i = k - 1}^{m - 2}\! {\left({{A^{i - m + k}}{{({A^i})^\top}} + {A^i}{{( {{A^{i - m + k}}})^\top}}} \right)}\sigma _\mathrm{p}^2,\!\!~2k \!-\! m \!\ge\! 1\\
\!\!\sum\limits_{i = 0}^{k - 2} \!{\left({{A^i}{{(\! {{A^{i + m - k}}})^\top}}  + {A^{i + m - k}}{{( {{A^i}})^\top}}} \right)\sigma _\mathrm{p}^2},\hspace{0.31cm}2k \!-\! m \!<\! 1
\end{array} \right.\nonumber\\
&\Theta_{(k,m)} \\
&\!\triangleq\! \left\{ \begin{array}{l}
\!\!\!\sum\limits_{i = 0}^{2k - m - 2} \!\!\!{\left({{A^i}{{\left( {{A^{i + m - k}}} \right)\!^\top}} \!\!+\! {A^{i + m - k}}{{\left( {{A^i}} \right)\!^\top}}} \right)\!\sigma _{\rm{p}}^2,2k \!-\! m \!\ge\! 2} \\
\!\!\!{\mathbf{O}_n}, \hspace{5.80cm}2k \!-\! m \!<\! 2
\end{array} \right. \nonumber\\
&{\Phi_{(k,m)}} \!\triangleq\! \left( {{A^{k - 1}} \!-\! {A^{m - 1}}} \right)\!{\left( {{A^{k - 1}} \!-\! {A^{m - 1}}} \right)\!^\top}\!\sigma^{2}_{\mathrm{i}} + 2\sigma _\mathrm{o}^2\mathbf{I}_{n} \nonumber\\
&\hspace{1.2cm} \!+\! \sum\limits_{i = k - 1}^{m - 2} \!\!\!{{A^i}} {\left( {{A^i}} \right)\!^\top}(\sigma_{\mathrm{p}}^2\!+\!\sigma^{2}_{\mathrm{a}}) \!+\! 2\sum\limits_{i = 0}^{k - 2}\! {{A^i}}{\left( {{A^i}} \right)\!^\top}\sigma_{\mathrm{p}}^2.
\end{align}\label{cov}
\end{subequations}
With the consideration of \eqref{kop3zd}, we can obtain from \eqref{ddvs} that
\begin{align}
\mathbf{E}[{{X_{(k,p)}}X_{(k,p)}^\top}] &= \sum\limits_{r = k}^{p - 1} {\sum\limits_{q = \mathbb{T}_r(1)}^{\mathbb{T}_r(\left| {{\mathbb{T}_r}} \right|)} {( {\Phi_{(r,q)} - \Omega_{(r,q)} - \Theta_{(r,q)}})} }\nonumber\\
& \triangleq {\Gamma _{(k,p)}} \triangleq M_{(k,p)}^{ - 2},\label{kop3p}
\end{align}
based on which, we define
\begin{subequations}
\begin{align}
&{\Upsilon_{(k,p)}} \triangleq \text{diag}\!\left\{\! {M_{(k,p)}},\ldots, {M_{(k,p)}} \!\right\} \!\in\! \mathbb{R}^{n{\mathfrak{n}_{(k,p)}} \times {n\mathfrak{n}_{(k,p)}}},\label{gh9aa}\\
&{\mathfrak{n}_{( {k,p})}} \triangleq \sum\limits_{q = k}^{p - 1} {( {( {q + 1})\left| {{\mathbb{T}_q}} \right| + \sum\limits_{j = 1}^{\left| {{\mathbb{T}_q}} \right|} {\left( {{\mathbb{T}_q}( j ) - q} \right)} })}.\label{gh9bb}
\end{align}\label{gh9}
\end{subequations}
In addition, the covariance matrix of random vector $\eta_{( {k,p})}$ defined by \eqref{ghb} is obtained as
\begin{align}
\mathcal{C}_{\mathrm{v}} &\triangleq  {\bf{E}}[\eta_{( {k,p})}\eta^{\top}_{( {k,p})}] = \left[ {\begin{array}{*{20}{c}}
{{\mathcal{C}_{xx}}}&\mathbf{O}&\mathbf{O}&\mathbf{O}\\
*&{\mathcal{C}_{ww}}&\mathbf{O}&\mathbf{O}\\
*&*&{\mathcal{C}_{\breve{f}\breve{f}}}&{{\mathcal{C}_{\breve{f}\widetilde{f}}}}\\
*&*&*&{{\mathcal{C}_{\widetilde{f} \widetilde {f}}}}
\end{array}} \right],\label{kop2}
\end{align}
where $\mathbf{O}$ denotes zero matrix with compatible dimensions, ${\mathcal{C}_{xx}} \triangleq \sigma_{\mathrm{i}}^2{\mathbf{I}_{\sum\limits_{m = k}^{p - 1} {\left| {{\mathbb{T}_m}} \right|} n}}$, ${\mathcal{C}_{ww}}$, $\mathcal{C}_{\breve{f}\breve{f}}$, ${\mathcal{C}_{\widetilde{f} \widetilde {f}}}$ and ${\mathcal{C}_{\breve{f}\widetilde{f}}}$ are given in subsections A--D of Appendix E, respectively.

Corresponding to \eqref{ghbvector} with \eqref{ghb}, we define a stacked matrix:
\begin{align}
{\Pi _{( {k,p})}} \triangleq [ {{\widehat{\mathcal{A}}_{(k,p)}}, ~~\mathbf{I}_{\sum\limits_{m = k}^{p - 1} \!\!{\left| {{\mathbb{T}_m}} \right|n}},~~{\breve{\mathcal{A}}_{\left( {k,p} \right)}},~~{\widetilde{\mathcal{A}}_{\left( {k,p} \right)}}} ],\label{ghmatrix}
\end{align}
where we define:
\begin{subequations}
\begin{align}
&A_{(k,m)} \triangleq A^{k-1} - A^{m-1}, \label{gh1}\\
&{\widehat A}_{k} \!\triangleq\!  \begin{cases}
\!{\text{diag} \!\left\{ {{A_{(k,\mathbb{T}_k(1))}}, \ldots ,{A_{(k,\mathbb{T}_k(\left| {{\mathbb{T}_k}} \right|))}}} \right\}}, \!\!\!\!\!&\text{if}~\mathbb{T}_k \!\neq\! \varnothing\\
\!`\mathrm{null}',\!\!\!\!\!&\text{otherwise}
\end{cases}\label{gh2}\\
&{\widehat{\mathcal{A}}_{(k,p)}} \triangleq \text{diag}\{ {{{\widehat A}_k},\;{{\widehat A}_{k + 1}},\; \ldots ,\;{{\widehat A}_{p - 2}},\;{{\widehat A}_{p - 1}}}\}, \label{gh3}\\
&{\breve{A}_k} \triangleq [{\mathbf{I}_{n},~A,~{A^2},~\ldots,~{A^{k - 2}}}],\label{gh4}\\
&{\breve{\bar{A}}}_{k} \triangleq  \begin{cases}
\text{diag}\{ {\underbrace{{{\breve{A}}_k}}_{1},\;\underbrace{{{\breve{A}}_{k}}}_{2},\; \ldots,\;{\underbrace{{\breve{A}}_{k}}_{\left| {{\mathbb{T}_k}} \right|}}}\}, \!\!&\text{if}~\mathbb{T}_k \!\neq\! \varnothing\\
`\mathrm{null}',\!\!&\text{otherwise}
\end{cases}\label{gh6}\\
&{\breve{\mathcal{A}}_{( {k,p} )}} \triangleq \text{diag}\{ {{\breve{\bar{A}}_k},~{\breve{\bar{A}}_{k + 1}},~ \ldots,~{\breve{\bar{A}}_{p - 2}},~{\breve{\bar{A}}_{p - 1}}}\},\label{gh5}\\
&{\widetilde{A}}^{m}_{k} \triangleq  [ {{-A^{k - 1}},~{-A^k},~\ldots,~{-A^{m - 3}},~{-A^{m - 2}}}],\label{gh5bba}\\
&{\widetilde{A}}_{k} \triangleq  \begin{cases}
\text{diag}\{ {{\widetilde{A}}^{\mathbb{T}_k(1)}_{k},~\ldots,~{\widetilde{A}}^{\mathbb{T}_k(\left| {{\mathbb{T}_k}} \right|)}_{k}}\}, &\text{if}~\mathbb{T}_k \!\neq\! \varnothing\\
`\mathrm{null}',&\text{otherwise}
\end{cases}\label{gh6}\\
&{\widetilde{\mathcal{A}}_{( {k,p})}} \triangleq \text{diag}\{ {{\widetilde{A}_k},~{\widetilde{A}_{k + 1}},~ \ldots,~{\widetilde{A}_{p - 2}},~{\widetilde{A}_{p - 1}}}\}.\label{gh7}
\end{align}\label{gh}
\end{subequations}
Finally, we define:
\begin{subequations}
\begin{align}
\underline{\mathcal{C}} &\!\triangleq\! \frac{\sigma_{\mathrm{o}}^2{\mathbf{I}_n}}{p-k} \!+\! \frac{{\sigma_{\mathrm{i}}^2}}{{{{(p - k)^2}}}}( {\sum\limits_{m = k}^{p - 1} {{A^{m - 1}}} }){( {\sum\limits_{m = k}^{p - 1} {{A^{m - 1}}} })^\top} \nonumber\\
&\hspace{1.2cm} \!+\! \frac{{\sigma_{\mathrm{p}}^2}}{{{{(p - k)^2}}}}\sum\limits_{m = 1}^{p - 2} {( {\sum\limits_{i = k - 1 - m}^{p - 2 - m}\!\!\! {\mathfrak{u}(i){A^i}} }){{( {\sum\limits_{i = k - 1 - m}^{p - 2 - m}\!\!\! {\mathfrak{u}(i){A^i}} })^\top}}} \nonumber\\
&\hspace{1.2cm} \!+\! \frac{{\sigma_{\mathrm{a}}^2}}{{{{(p - k)^2}}}}( {\sum\limits_{m = k}^{p - 1} {\sum\limits_{i = 0}^{m - 2} {{A^i}} } }){( {\sum\limits_{m = k}^{p - 1} {\sum\limits_{i = 0}^{m - 2} {{A^i}} } })^\top},\label{kmcf3} \\
\overline{\mathcal{C}} &\!\triangleq\! \frac{{( {\sigma_{\mathrm{p}}^2 \!+\! \sigma _{\rm{o}}^2}){{\bf{I}}_n} \!+\! \sigma _{\rm{o}}^2A{A^ \top }}}{{p - k}} \!-\! \frac{{( {p \!-\! k \!-\! 1})\sigma _{\rm{o}}^2(A \!+\! {A^ \top })}}{{{{( {p \!-\! k})^2}}}},\label{kmcf1ab}\\
\underline{\mathfrak{h}} &\!\triangleq\!  \frac{{n( {\sigma _{\rm{p}}^2 \!+\! \sigma _{\rm{o}}^2}) \!+\! ||A||_{\rm{F}}^2\sigma _{\rm{o}}^2}}{{p - k}} \!-\! \frac{{{\rm{2}}( {p \!-\! k \!-\! 1})\sigma _{\rm{o}}^2{\rm{tr}}(A)}}{{{{( {p \!-\! k})}^2}}},\label{kmcfa2}\\
\overline{\mathfrak{h}} &\!\triangleq\! \frac{{\sigma_{\mathrm{i}}^2}}{{(p - k)^{2}}}||\!{\sum\limits_{m = k}^{p - 1} {{A^{m - 1}}} }||_{\mathrm{F}}^2 \!+\! \frac{{\sigma_{\mathrm{a}}^2}}{(p - k)^{2}}||\!{\sum\limits_{m = k}^{p - 1} {\sum\limits_{i = 0}^{m - 2}\!\! {{A^i}} } }||_{\mathrm{F}}^2  \nonumber\\
&\hspace{0.95cm} \!+\! \frac{{\sigma_{\mathrm{p}}^2}}{(p - k)^{2}}\sum\limits_{m = 1}^{p - 2} {||{\sum\limits_{i = k - 1 - m}^{p - 2 - m}\!\!\! {\mathfrak{u}(i){A^i}} }||_{\mathrm{F}}^2} + \frac{n\sigma_{\mathrm{o}}^2}{p-k}, \label{kmcf2}
\end{align}\label{mkooo}
\end{subequations}
\!\!where $\mathfrak{u}(\cdot)$ denotes the unit step fruntion, i.e., $\mathfrak{u}(t) = 1$ if $t > 0$ and $\mathfrak{u}(t) = 0$, otherwise.

\subsection{Derived Sample Complexity}
\vspace{-0.1cm}
With the definitions \eqref{ddvs}, \eqref{kop3p}--\eqref{ghmatrix} and \eqref{mkooo} at hand, we present an auxiliary proposition, which is used in studying the sample complexity of the proposed model inference procedure.
\begin{prop}
Under Assumptions \ref{asprv}-1) and \ref{asprv}-2),  we have \eqref{hhhccc3}--\eqref{hhhccc1} for $\varepsilon \in [0, \frac{1}{2})$ and some universal constant $\mathfrak{c} > 0$.
\begin{figure*}
\begin{align}
&\mathbf{P}\!\left[ {\left| {{{\left\| {\sum\limits_{m = k}^{p - 1} {\frac{{{\bf{r}}(m)}}{{p - k}}}  - \mu {{\bf{1}}_n}} \right\|}^2} - \overline{\mathfrak{h}}} \right| > {\rho _1}} \right] \le 2\cdot{e^{{\frac{-1}{{\mathfrak{c}{\kappa ^2}}}\min \{ {\frac{{\rho _1^2}}{{n\underline{\mathcal{C}}}},~~{\rho _1}}\}}}}, \label{hhhccc3}\\
&\mathbf{P}\!\left[ {\left| {{{\left\| {\frac{1}{{p - k}}\sum\limits_{m = k}^{p - 1}  ({\bf{f}}(m) + {\bf{w}}(m + 1) - A{\bf{w}}(m)) - ( {\mu {{\bf{1}}_n} - A\mu {{\bf{1}}_n}} )} \right\|^2}} - \underline{\mathfrak{h}}} \right| > {\rho _2}} \right] \le 2\cdot{e^{{ - \frac{1}{{\mathfrak{c}{\kappa^2}}}\min \left\{ {\frac{{\rho _2^2}}{{n\overline{\mathcal{C}}}},~~{\rho _2}} \right\}}}}, \label{hhhccc2}\\
&\mathbf{P}\!\left[ {\left\| {M_{(k,p)}^ \top {X_{(k,p)}}X_{(k,p)}^ \top {M_{(k,p)}} - {{\bf{I}}_n}} \right\| > \rho_{3} } \right]
\leq 2 \cdot {( {\frac{2}{\varepsilon } + 1})^n} \cdot e^{-\frac{1}{{\mathfrak{c}{\kappa^2}}} \min \left\{\! {\frac{{{(1 - 2\varepsilon )^2\rho_{3} ^2}}}{{{{{\mathfrak{n}_{({k},{p})}||{{\Pi}^{\top}_{({{k},{p}})}}{{\Upsilon}_{({k},{p})}}||^{4}}}}|| {{{\mathcal{C}}_\mathrm{v}}} ||}},~~\frac{{(1 - 2\varepsilon)\rho_{3}}}{{{||{\Pi}^{\top}_{({{k},{p}})}\!{\Upsilon}_{({k},{p})}} ||^{2}}}} \!\right\}}. \label{hhhccc1}
\end{align}
\end{figure*}\label{ptopk1a}
\end{prop}
\begin{IEEEproof}
See Appendix F.
\end{IEEEproof}

Leveraging Proposition \ref{ptopk1a}, the sample complexity is presented in the following theorem.
\begin{thm} Consider the inference solution \eqref{selffn}. For any $\varepsilon \in [0, \frac{1}{2})$, and any $\rho_{1},\rho_{2},\rho_{3} \in (0, 1)$, and any $\phi > 0$, and
 \begin{align}
\widehat{\phi}  = \left(\! {\sqrt {{\rho_1} + \overline{\mathfrak{h}}}  + \sqrt n \mu }\!\right)\!\phi  + {\sqrt {{\rho_2} + \underline{\mathfrak{h}}}  + \left\| \mu {\mathbf{1}_n} - A\mu {\mathbf{1}_n}  \right\|}, \nonumber
\end{align}we have:
\begin{enumerate}
  \item $\mathbf{P} (||A-A_{\mathrm{if}}|| \leq \phi) \geq 1 - \delta$, as long as the following conditions hold:
  \begin{align}
&\min \left\{ {\frac{{{(1 - 2\varepsilon )^2\rho_{3}^2}}}{{{{{\mathfrak{n}_{({k},{p})}||{{\Pi}^{\top}_{({{k},{p}})}}\!{{\Upsilon}_{({k},{p})}}||^{4}}}}|| {{{\mathcal{C}}_\mathrm{v}}} ||}},~\frac{{(1 - 2\varepsilon)\rho_{3}}}{{{||{\Pi}^{\top}_{({{k},{p}})}{\Upsilon}_{({k},{p})}} ||^{2}}}} \right\} \nonumber\\
&\ge \frac{{\gamma ^2}}{2}\!\ln\! \frac{{4 {( {\frac{2}{\varepsilon } + 1})^n}}}{\delta },\label{fg0aa}\\
&{\lambda _{\min }}\!\!\left( {{\Gamma _{(k,p)}}} \!\right) \!\ge\! \frac{{32\mathfrak{c}{\kappa ^2}}}{{{\phi ^2}\left( {1 \!-\! \rho_{3} } \right)}}\!\ln \!\!\left(\!\!\! {{{\left( {\frac{{1 \!-\! \rho_{3} }}{2}} \right)^{0.5n}}}\frac{{2 \cdot {5^n}}}{\delta }} \!\right)\!.\label{fg4a}
\end{align}
  \item $\mathbf{P}[|| {{\alpha_{\text{if}}} - \alpha }|| \leq \widehat{\phi}] \geq 4\cdot(1 - \delta)\cdot{e^{{ - \frac{\min \{ {\frac{{\rho _2^2}}{{n\overline{\mathcal{C}}}},{\rho _2}}\}+{\min \{ {\frac{{\rho _1^2}}{{n\underline{\mathcal{C}}}},{\rho _1}}\}}}{{\mathfrak{c}{\kappa^2}}}}}}$, as long as \eqref{fg0aa} and \eqref{fg4a} hold.
\end{enumerate}
\label{ptopk1}
\end{thm}

\begin{IEEEproof}
See Appendix G.
\end{IEEEproof}

\begin{rem}There exists an optimal $\varepsilon$ for the condition \eqref{fg0aa}. Let us define $g( \varepsilon) \triangleq {(1 - 2\varepsilon )^2}a - b\ln \left( {{{(\frac{2}{\varepsilon } + 1)}^n}c} \right)$, with $a = \frac{{{\rho_{3}^2}}}{{{{{\mathfrak{n}_{({k},{p})}||{{\Pi}^{\top}_{({{k},{p}})}}\!{{\Upsilon}_{({k},{p})}}||^{4}}}}|| {{{\mathcal{C}}_\mathrm{v}}} ||}},b = \frac{{{\gamma ^2}}}{2},c = \frac{4}{\delta }$. Given $\frac{{{(1 - 2\varepsilon )^2\rho_{3} ^2}}}{{{{{\mathfrak{n}_{({k},{p})}||{{\Pi}^{\top}_{({{k},{p}})}}\!{{\Upsilon}_{({k},{p})}}||^{4}}}}|| {{{\mathcal{C}}_\mathrm{v}}} ||}} \le \frac{{(1 \!-\! 2\varepsilon)\rho_{3}}}{{{||{\Pi}^{\top}_{({{k},{p}})}{\Upsilon}_{({k},{p})}} ||^{2}}}$, the optimal parameter can be numerically solved from the relation $(1 - 2{\varepsilon _{\mathrm{opt}}})(2 + {\varepsilon _{\mathrm{opt}}}){\varepsilon _{\mathrm{opt}}} = \frac{{nb}}{{2a}}$ with $\varepsilon _{\mathrm{opt}} \in [0, \frac{1}{2})$, which is obtained through solving $\frac{{\mathrm{d}g\left( \varepsilon  \right)}}{{\mathrm{d}\varepsilon }} = 0$.
\end{rem}

\section{Sample-Complexity Bounds}
\vspace{-0.1cm}
Building on the obtained sample complexity, the upper bound on the observation numbers, the lower bound on the model error, and the derived Redundant Data Processor are investigated in this section. We note that all of the expected results rely on the inequalities \eqref{fg0aa} and \eqref{fg4a}, which however cannot be used in the current form due to the unknown $A$ included in ${{\Pi}^{\top}_{({{k},{p}})}}\!{{\Upsilon}_{({k},{p})}}$ and $\Gamma _{(k,p)}$. As a remedy, we first provide two estimation bounds pertaining to $\Gamma _{(k,p)}$ and ${{\Pi}^{\top}_{({{k},{p}})}}\!{{\Upsilon}_{({k},{p})}}$.

\vspace{-0.2cm}
\subsection{Bounds on $\Gamma _{(k,p)}$ and ${{\Pi}^{\top}_{({{k},{p}})}}\!{{\Upsilon}_{({k},{p})}}$}
We let $\sigma_{i}\left(A\right)$ denote the $i$th singular value of matrix $A$. We assume the following bounds pertaining to $A$ are known:
\begin{subequations}
\begin{align}
\widehat{\underline{\sigma}}_{A} &\le \mathop {\min }\limits_{i \in \{ 1, \ldots ,n\}}\!\! \left\{ {{\sigma _i}\!\left( A \right)} \right\},\label{hx1}\\
\widetilde{\underline{\sigma}}_{A} &\le \mathop {\min }\limits_{m \in \left\{ {k + 1, \ldots ,p} \right\},i \in \left\{ {1, \ldots ,n} \right\}} \left\{ {{{\sigma _i( {A^{m - k} - \mathbf{I}_{n}} )}}} \right\},\label{hx2}\\
\overline{\widetilde{\sigma}}_{A} &\geq \left\| A \right\|,\label{hx3}\\
{\overline{\widehat{\sigma}}}_{A} &\geq \mathop {\max }\limits_{m \in \left\{ {k + 1, \ldots ,p} \right\}} \left\{ {{{\left\| {{A^{k - 1}} - {A^{m - 1}}} \right\|}}} \right\},\label{hx4}
\end{align}
\end{subequations}
based on which, we define the following functions:
\begin{subequations}
\begin{align}
\mathfrak{f}_{1}(A) &\triangleq \widehat{\underline{\sigma}}_{A}^{2k - 2}\widetilde{\underline{\sigma}}_{A}^{2}\sigma_{\mathrm{i}}^2 + \widehat{\underline{\sigma}}_{A}^{2k-2}\sigma_{\rm{p}}^2 + 2\sigma_{\mathrm{o}}^2,\label{defho}\\
\mathfrak{f}_{2}(A) &\triangleq \widehat{\underline{\sigma}}_{A}^{2k - 2}\widetilde{\underline{\sigma}}_{A}^{2} \sigma_{\mathrm{i}}^2 + 2\sigma_{\mathrm{o}}^2.\label{defho2}
\end{align}
\end{subequations}
With the definitions at hand, we present a proposition, leveraging which we can derive the lower bound on ${\lambda _{\min }}\left({{\Gamma _{(k,p)}}} \right)$.
\begin{prop}
Consider the dynamics \eqref{dv1}. Under Assumption \ref{asprv}-1), we have
\begin{align}
{\bf{E}}\!\left[\mathbf{r}_k^m(\mathbf{r}_k^m)^{\top}\right] \geq \begin{cases}
\mathfrak{f}_{1}(A)\mathbf{I}_{n}, &\text{if}~m > 2k- 1\\
\mathfrak{f}_{2}(A)\mathbf{I}_{n},&\text{if}~m  \leq 2k- 1
\end{cases}.\label{ckj}
\end{align}
\label{pop2}
\end{prop}
\begin{IEEEproof}
See Appendix H.
\end{IEEEproof}

Given the observation data in a time interval $\{k, k+1, \ldots, p\}$, the proposed model inference needs $p-k+1$ basis data. It follows from \eqref{defho} and \eqref{defho2} that $\mathfrak{f}_{2}(A) \leq \mathfrak{f}_{1}(A)$, which together with \eqref{ddvs}, \eqref{kop3p} and \eqref{ckj} imply that
\begin{align}
{\lambda _{\min }}\left({{\Gamma _{(k,p)}}} \right) \geq \mathfrak{f}_{2}(A)(p-k+1), \label{sclbad}
\end{align}
leveraging which, we next present the obtained upper bound on $|| {{\Pi}^{\top}_{({{k},{p}})}}\!{{\Upsilon}_{({k},{p})}} ||$ in the following proposition.
\begin{prop} For matrices ${{\Pi}_{({{k},{p}})}}$ and ${{\Upsilon}_{({k},{p})}}$ given by \eqref{ghmatrix} and \eqref{gh9}, respectively, the following inequality holds:
\begin{align}
|| {{\Pi}^{\top}_{({{k},{p}})}}\!{{\Upsilon}_{({k},{p})}} ||^2 \le \frac{{{{\mathfrak{g}}^{2}_{(k,p)}}}}{{\sum\limits_{r = k}^{p - 1} {\sum\limits_{q = {\mathbb{T}_r}(1)}^{{\mathbb{T}_r}\left( {\left| {{\mathbb{T}_r}} \right|} \right)} {{\mathfrak{f}_{( {r,q})}}} } }} \triangleq \mathfrak{p}_{(k,p)}, \label{aah}
\end{align}
where
\begin{align}
&\mathfrak{g}_{(k,p)} \triangleq 1 + \overline{\widehat{\sigma}} + \mathop {\max }\limits_{q \in \left\{ {k, \ldots ,p - 1} \right\}}\left\{ {\frac{{{\overline{\widetilde{\sigma}}_A} - \overline{\widetilde{\sigma}}_A^{q - 1}}}{{1 - {\overline{\widetilde{\sigma}}_A}}}}\right\} \nonumber\\
&\hspace{2.1cm}+ \mathop {\max }\limits_{j < m \in \left\{ {k, \ldots ,p - 1} \right\}}\left\{{\frac{{\overline{\widetilde{\sigma}}_A^{j - 1} -\overline{\widetilde{\sigma}}_A^{m - 1}}}{{1 - {\overline{\widetilde{\sigma}}_A}}}}\right\},\label{cvop} \\
&\mathfrak{f}_{(k,m)} \triangleq \begin{cases}
\mathfrak{f}_{1}(A), &\text{if}~m > 2k- 1\\
\mathfrak{f}_{2}(A), &\text{if}~m  \leq 2k- 1
\end{cases}.\label{ckjdef}
\end{align}
\label{pppt}
\end{prop}
\begin{IEEEproof}
See Appendix I.
\end{IEEEproof}

With the obtained upper bound $\mathfrak{p}_{(k,p)}$ \eqref{aah}, the condition \eqref{fg0aa} updates as
\begin{align}
\!\!\!\min\! \left\{\! {\frac{{{(1 \!-\! 2\varepsilon )^2\rho_{3}^2}}}{{\mathfrak{n}_{({k},{p})}\mathfrak{p}^{2}_{(k,p)}|| {{{\mathcal{C}}_\mathrm{v}}} ||}},\frac{{(1 \!-\! 2\varepsilon )\rho_{3}}}{{\mathfrak{p}_{(k,p)}}}} \!\right\} \!\ge\! \frac{{\gamma ^2}}{2}\!\ln\! \frac{{4 {( {\frac{2}{\varepsilon } \!+\! 1})^n}}}{\delta }.\label{fg0}
\end{align}

\subsection{Sample-Complexity Upper Bound}
\vspace{-0.1cm}
Theorem \ref{ptopk1} indicates that the vector-error bound $\widehat{\phi}$ and its confidence completely depend on the matrix-error bound $\phi$ and its associated confidence and conditions \eqref{fg0aa} and \eqref{fg4a}. This observation means that the upper bound on the number of observations (i.e., $p-k+1$) that is sufficient to achieve PAC is determined by the inference of system matrix, which now can be implicitly estimated from
\eqref{fg4a}. Using the lower bound relation \eqref{sclbad}, the sample-complexity upper bound on the number of observations $l \triangleq p - k + 1$ is obtained as
\vspace{-0.15cm}
\begin{align}
\!\!l_{\mathrm{up}}(\phi, \delta) \!=\! \frac{{32\mathfrak{c}{\kappa ^2}}}{{{\left( {1 \!-\! \rho_{3} } \right)\phi^2\mathfrak{f}_{2}(A)}}}\! \ln \!\!\left(\! {{{\left(\! {\frac{{1 \!-\! \rho_{3} }}{2}} \!\right)^{0.5n}}}\frac{{2 \cdot {5^n}}}{\delta }} \!\right) \!+\! 1, \label{sclb}
\end{align}
such that the matrix inference solution \eqref{self} is $\left(\theta, \delta\right)$--PAC for $p \geq l_{\mathrm{up}}(\phi, \delta) + k - 1$.

\begin{rem}
We note that the sample-complexity upper bound can be further reduced with more knowledge of real system matrix, such as $\mathop {\max}\limits_{i \in \{ 1, \ldots ,n\}}\!\! \left\{ {{\sigma _i}\!\left( A \right)} \right\} < 1$, as assumed in \cite{jedra2020finite,simchowitz2018learning,sarkar2019near,sarkar2019finite,oymak2019non,banerjee2019random, foster2020learning}. The relation \eqref{fg4a}, in conjunction with \eqref{cov} and \eqref{kop3p}, implies that the knowledge of system matrix can be completely dropped, which however results in larger upper bound especially when the variance of observation noise is small.
\end{rem}

\vspace{-0.5cm}
\subsection{Redundant Data Processor}
\vspace{-0.15cm}
As demonstrated by Figure \ref{sp} (b), given the basis data, the resulted redundant data has complex influence on model inference error. This observation motivates  to incorporate the Redundant Data Processor into our proposed model inference procedure, as show in Figure \ref{modelinfer} (a), which optimally uses basis and redundant data to guarantee $\left(\theta, \delta\right)$--PAC and reduce the model error. We let $\mathbb{V}_{k,p}$ denote the set of all basis and redundant vector data, i.e.,
\vspace{-0.2cm}
\begin{align}
\mathbb{V}_{(k,p)}  = \left\{ {\left. {\mathbf{r}_r^q} \right|q \!>\! r \!\in\! \left\{ {k, k + 1, \ldots ,p} \right\}} \right\}, \label{pd}
\end{align}
We let $\mathbb{I}_{(k,p)} \subseteq \mathbb{V}_{(k,p)}$ denote the set of basis and redundant vector data used by the inference computation \eqref{self}.

The sample complexity analysis presented in Theorem \ref{ptopk1} indicates that the choice of $\mathbb{I}_{(k,p)}$ directly influences the left-hand terms of \eqref{fg0} and \eqref{fg4a}, whose larger magnitudes are preferred, since the proposed inference solution is more likely to be $\left(\theta, \delta\right)$--PAC. The relation \eqref{kop3p} implies that the left-hand term of \eqref{fg4a}, i.e., ${\lambda _{\min }}\!\left( {{\Gamma _{(k,p)}}} \right)$, is strictly increasing with respect to the number of redundant vector data, which is due to ${\bf{E}}\left[\mathbf{r}_k^m(\mathbf{r}_k^m)^{\top}\right] > 0$, $m > k \in \mathbb{N}$. Then, observing \eqref{fg0} and \eqref{fg4a} we conclude that the left-hand term of \eqref{fg0} is more sensitive to redundant data. Motivated by this observation, we let the Redundant Data Processor leverage only the left-hand term of \eqref{fg0} to determine the usage of redundant and basis data, i.e.,
\vspace{-0.2cm}
\begin{align}
\!\!\!\mathbb{I}_{(k,p)} \!=\!\! \mathop {\arg \max }\limits_{{{\widetilde{\mathbb{I}}}_{( {k,p})}} \subseteq {\mathbb{V}_{( {k,p})}}} \!\!\!\left\{\!\min \!\left\{\! {\frac{{{(1 \!-\! 2\varepsilon )^2\rho_{3}^2}}}{{\mathfrak{n}_{({k},{p})}\mathfrak{p}^{2}_{(k,p)}|| {{{\mathcal{C}}_\mathrm{v}}} ||}},\frac{{(1 \!-\! 2\varepsilon )\rho_{3}}}{{\mathfrak{p}_{(k,p)}}}} \!\!\right\} \!\!\right\}\!\!.\label{rpd}
\end{align}

\vspace{-0.6cm}
\subsection{$\left(\theta, \delta\right)$--PAC Verification}
\vspace{-0.1cm}
We note that the computation of sample-complexity upper bound needs the prescribed levels of model accuracy $\phi$ and confidence $1 - \delta$, which however in turn relies on the estimated upper bound. This implies that if the prescribed $\phi$ and $\delta$ are not reasonable, the bound computation \eqref{sclb} is not feasible. To address this issue, we propose an algorithm of $\left(\theta, \delta\right)$--PAC verification and updating, as described by Algorithm~1.
\vspace{-0.2cm}
\begin{align}
\!\!\!l_{\mathrm{up}}(\phi, \delta) \!\geq\! \frac{{32c{\kappa ^2}}}{{{( {1 \!-\! \rho_{3} })\phi^2\mathfrak{f}_{2}(A)}}} \ln \!\!\left(\!\! {{{\left(\! {\frac{{1 \!-\! \rho_{3} }}{2}} \!\right)^{0.5n}}}\frac{{2 \cdot {5^n}}}{\delta }} \!\right) \!+\! 1. \label{rpdkkc}
\end{align}
\vspace{-0.6cm}
\begin{algorithm}\small
  \caption{$\left(\theta, \delta\right)$--PAC Verification and Updating}
  \KwIn{$k \in \mathbb{N}$,~ $0 < \rho_{3} < 1$,~ $0 < \delta < 1$,~ $\phi > 0$,~$0 \leq \varepsilon < \frac{1}{2}$.}
   Compute sample-complexity upper bound  $l_{\mathrm{up}}(\phi, \delta)$ according to \eqref{sclb};\\
   Choose basis and redundant vector data according to \eqref{rpd} with $p \geq k + l_{\mathrm{up}}(\phi, \delta) - 1$;\\
    \eIf{condition \eqref{fg0} does not hold}
    {\eIf{conditions \eqref{fg0} and \eqref{rpdkkc} hold through adjusting $\rho_{3}$ and $\varepsilon$}
    {Adjust $\rho_{3}$ and $\varepsilon$ such that \eqref{fg0} and \eqref{rpdkkc} hold;\\
    Go to Step 1;\\
    }
    {Adjust $\delta$ or/and $\phi$;\\
    Go to Step 1;\\}}
    {Break.}
\end{algorithm}
\vspace{-0.5cm}
\begin{rem}
Lines 4-9 of Algorithm 1 indicate that if the conditions \eqref{fg0} and \eqref{rpdkkc} cannot hold through adjusting $\rho_{3}$ and $\varepsilon$, we have to reduce PAC for the feasibility of \eqref{fg0}.
\end{rem}
\vspace{-0.5cm}
\subsection{Sample-Complexity Lower Bound}
\vspace{-0.1cm}
We note that the condition \eqref{fg4a} is equivalent to
\vspace{-0.1cm}
\begin{align}
\phi \!\ge\! \sqrt {\frac{{32c{\kappa ^2}}}{{{\lambda _{\min }}\left( {{\Gamma _{(k,p)}}} \right)\!\left( {1 - \rho_{3} } \right)}}\ln\!\left(\! {{{\left( {\frac{{1 - \rho_{3} }}{2}} \right)}^{0.5n}}\frac{{2 \cdot {5^n}}}{\delta }} \right)} \!\triangleq\! \mathfrak{m}, \nonumber
\end{align}
which indicates that the sample-complexity lower bound on matrix inference error can be obtained through estimating the upper bound on $\mathfrak{m}$. Considering \eqref{sclbad}, the lower bound is then obtained as
\vspace{-0.1cm}
\begin{align}
\mathfrak{m}_{\mathrm{lo}}(\phi, \delta) \!=\! \sqrt {\frac{{32c{\kappa ^2}}}{{{\mathfrak{f}_{2}(A)l_{\mathrm{up}}(\phi, \delta)}\left( {1 - \rho_{3} } \right)}}\!\ln \!\!\left(\!\! {{{\left(\! {\frac{{1 - \rho_{3} }}{2}} \!\right)}^{0.5n}}\frac{{2 \cdot {5^n}}}{\delta }} \!\right)},\nonumber
\end{align}
such that \eqref{fg4a} holds if $\phi \geq \mathfrak{m}_{\mathrm{lo}}(\phi, \delta)$.

\vspace{-0.1cm}
\section{Simulation}
\vspace{-0.1cm}
\subsection{Data Processor}
\vspace{-0.1cm}
In \cite{jedra2020finite,simchowitz2018learning,sarkar2019near,sarkar2019finite,oymak2019non,banerjee2019random}, using raw data and assuming $\mathbf{a} = \mathbf{0}_{n}$, system matrix is identified via ordinary least-square estimator as
\vspace{-0.1cm}
\begin{align}
A_{\mathrm{os}} = YW^\top{( {WW^{\top}})^{ - 1}}, \label{ols}
\end{align}
where $W \!=\! [{\bf{r}}{(k)}, \ldots, {\bf{r}}{(p-1)}]$, $Y \!=\! [{\bf{r}}{(k+1)},\ldots,{\bf{r}}{(p)}]$.

We use the mean metrics
$\mathbf{E}\left[ {\left\| {{A_{\mathbf{if}}} - A} \right\|} \right]$ and $\mathbf{E}\left[ {\left\| {{A_{\mathbf{os}}} - A} \right\|} \right]$
to quantify the influence of observation noise on the model errors of the proposed model inference  \eqref{self} (without using the redundant data) and the direct ordinary least-square estimator \eqref{ols} (without processing the observation data), respectively. To perform the comparison, we consider a system \eqref{realdyna} with
\begin{align}
A = \alpha\left[{\begin{array}{*{20}{c}}
{1}&{0}&{0}&{1}\\
{0}&{1}&{0}&{1}\\
{1}&{0}&{1}&{0}\\
{1}&{0}&{1}&{1}
\end{array}} \right], ~~\mathbf{a} = \mathbf{0}_4. \label{sysma}
\end{align}
$[\mathbf{f}(k)]_{i} \iidsim \mathcal{U}( - 1,1)$, $[\mathbf{w}(k)]_{i} \iidsim \mathcal{U}(0,1)$, $[\mathbf{x}(1)]_{i} \iidsim \mathcal{U}(-1,1)$, $ i \in \{1,2,3,4\}$, where $\mathcal{U}(\cdot)$ denote a uniform distribution. Given a time interval $\{k, k+1, \ldots, p\}$, the proposed model inference uses only the basis data: $r^{m+1}_{m}, m = k, k+1, \ldots, p-1$.  We let $\alpha = 0.5$. With 20000 random samples, the defined metrics under different observation number, i.e., $p-k+1$, are shown in Figure \ref{sp} (a), which demonstrates the promising advantage of incorporating Data Processor from the perspective of reducing model error.

\vspace{-0.4cm}
\subsection{Redundant Vector Data}
\begin{figure}
\centering{
\includegraphics[scale=0.555]{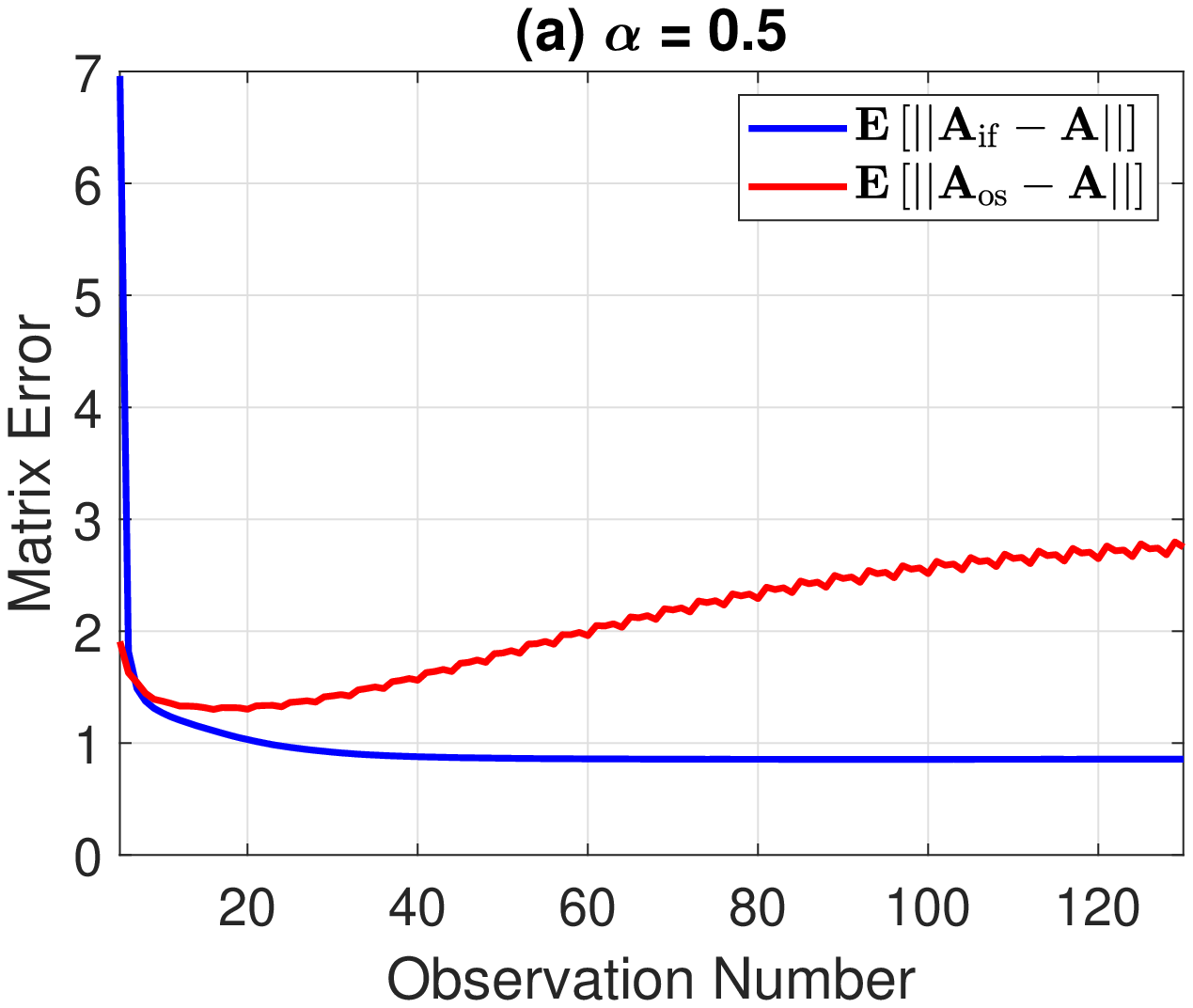}\\
\includegraphics[scale=0.415]{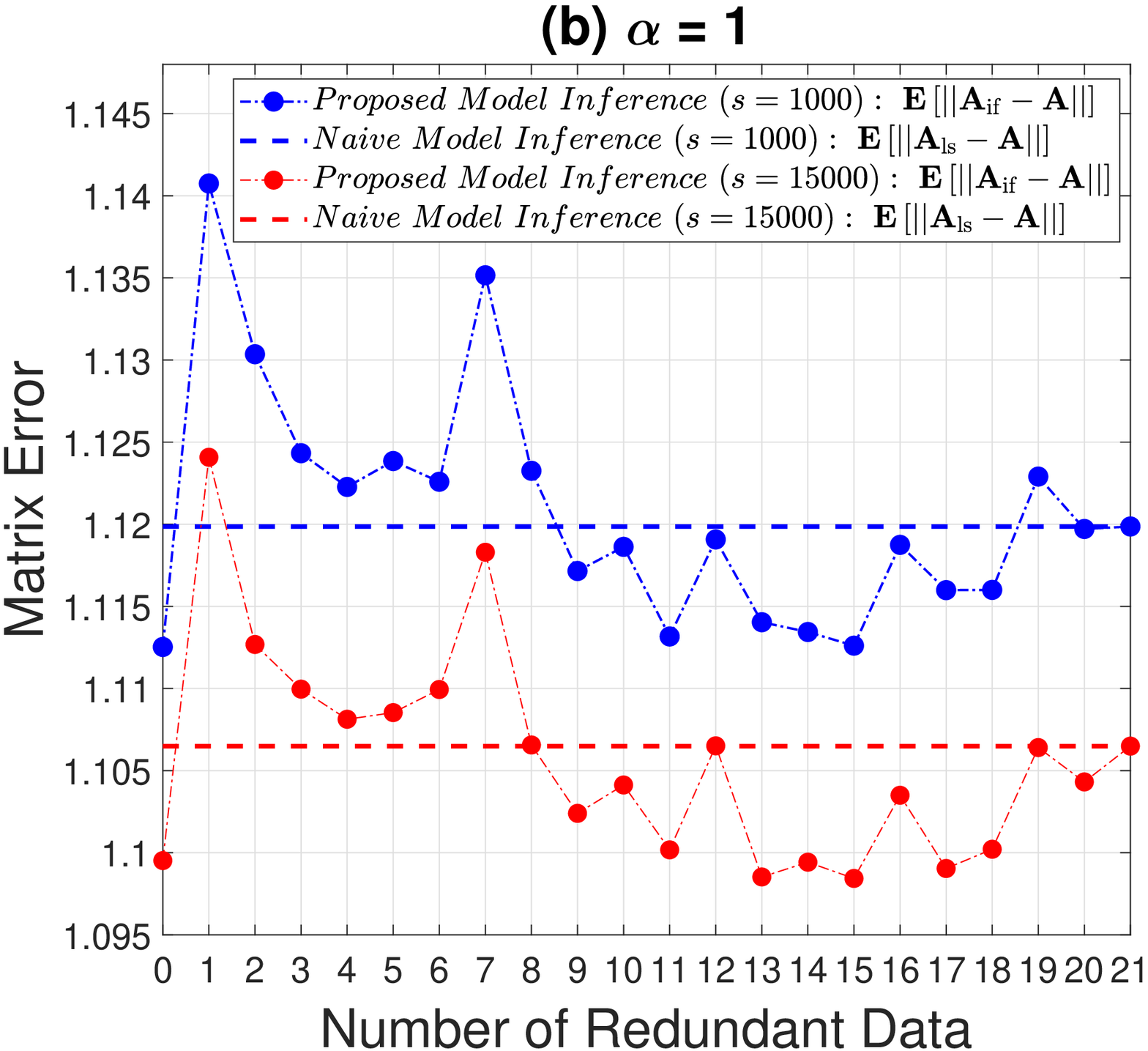}}
\caption{(a) Proposed model inference is much more accurate, (b) the mean of matrix error.}
\label{sp}
\end{figure}

\vspace{-0.1cm}
We let $\alpha =  1$, the initial condition, process and observation noise follow uniform distribution:
$[\mathbf{x}(1)]_{i} \iidsim \mathcal{U}(-1,1)$, $[\mathbf{f}(k)]_{i} \iidsim \mathcal{U}(-10,10)$ and $[\mathbf{w}(k)]_{i} \iidsim \mathcal{U}(0,2)$, $i = 1, 2, 3, 4$. We fix the observation-starting time and data length as $k = 1$ and $ l = 8$, respectively. We let $\breve{X}_{1}$ include the considered seven basis vector data:
$\mathbf{r}^{2}_{1}$, $\mathbf{r}^{3}_{1}$, \ldots, $\mathbf{r}^{8}_{1}$, based on which, we obtain twenty-one redundant vector data: $\mathbf{r}^{3}_{2}$, $\mathbf{r}^{4}_{2}$, \ldots, $\mathbf{r}^{8}_{2}$, $\mathbf{r}^{4}_{3}$, $\mathbf{r}^{5}_{3}$, \ldots, $\mathbf{r}^{8}_{3}$,$\mathbf{r}^{5}_{4}$, $\mathbf{r}^{6}_{4}$, \ldots, $\mathbf{r}^{8}_{4}$,$\mathbf{r}^{6}_{5}$, $\mathbf{r}^{7}_{5}$, $\mathbf{r}^{8}_{5}$, $\mathbf{r}^{7}_{6}$, $\mathbf{r}^{8}_{6}$, $\mathbf{r}^{8}_{7}$,
increasingly using which, the means of matrix error $\|A_{\mathrm{if}} - A\|$ with different numbers of random samples $s$ are shown in Figure \ref{sp} (b), which shows:
\begin{itemize}
  \item redundant data has significant influence on model error, which highlights the importance of Redundant Data Processor in model inference;
  \item incorporating twenty-one redundant vectors into the proposed inference procedure such that \eqref{npp} holds, the naive and proposed inference procedures have the same model error, which demonstrates statement 2) in Theorem \ref{fthmp}.
\end{itemize}

\vspace{-0.5cm}
\subsection{Sample Complexity}
\begin{figure}
\centering{
\includegraphics[scale=0.37]{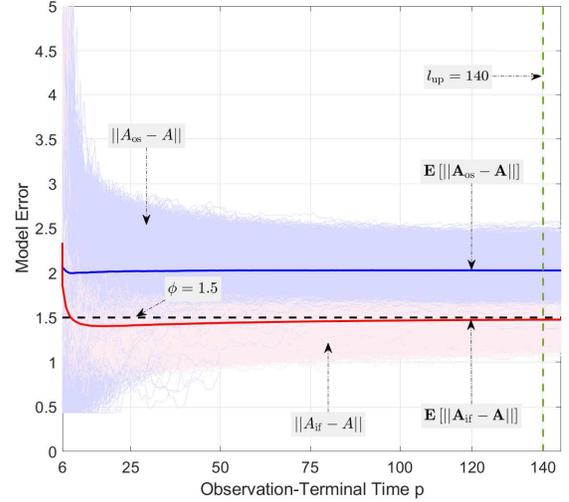}}
\caption{With fixed $k=1$, model errors under different observation-terminal time: $(\phi = 1.5, \delta = 0.2811)$-PAC is achieved and $\mathbf{E}[\|A_{\mathrm{if}} - A\|]$ is much smaller than $\mathbf{E}[\|A_{\mathrm{os}} - A\|]$.}
\label{spsp}
\end{figure}

To demonstrate the advantage of the proposed model inference in mitigating the influence of the observation noise in the model error, as well as the obtained sample-complexity upper bound, we let $[\mathbf{w}(k)]_{i} \iidsim \mathcal{U}(0,1)$,  $[\mathbf{w}(k)]_{i} \iidsim \mathcal{U}(-0.05,0.05)$, $[\mathbf{x}(1)]_{i} \iidsim \mathcal{U}(-0.05,0.05)$, $i = 1, \ldots, 4$. With the knowledge of distributions, we can set $\gamma = 0.0833$, $\kappa = 2.2\sqrt{2}$, $\mathfrak{c} = 9.5$. We consider an unstable real matrix \eqref{sysma} with $\alpha = 0.44$. We assume we only know its upper bound on singular values: $\widehat{\overline{\sigma}} \left( A \right) = 1.004$. We let the observation-starting time be the initial time. By Algorithm~1, we can set $(\phi = 1.5, \delta = 0.2811)$-PAC, and we obtain an upper bound on the number of observations as $l_{\mathrm{up}} = 140$, and set the observation terminal time as $p = 145$. The matrix errors and their means with 10000 samples under different terminal time are shown in Figure \ref{spsp}, from which we observe that 1) the model error of the proposed model inference is much smaller than that of the ordinary least-square estimator \eqref{ols}, and 2) $(\phi = 1.5, \delta = 0.2811)$-PAC is achieved when $p > 140$.

\vspace{-0.1cm}
\section{Conclusion}
\vspace{-0.1cm}
This paper has studied the problem of model inference using finite-time noisy state data from a single system trajectory. We have systematically compared the proposed model inference and the naive model inference and highlighted the advantages of the proposed inference procedure with incorporation of the redundant data processor. We then investigated the bounds on sample complexity of the proposed model inference in the presence of observation noise which leads to the dependence of samples on time and coordinates. The effectiveness of the proposed model inference procedures have been numerically demonstrated.

Our analysis suggests two future research directions: 1) finite-time model inference of continuous-time systems via sampling, with well-calibrated state measurements, i.e., neither underestimation nor overestimation, 2) and sample complexity of the model inference in the presence of process and observation noise with time-varying means and dependencies.

\vspace{-0.1cm}
\appendices
\section*{Appendix A: Auxiliary Lemmas}
We present the auxiliary lemmas for the proofs.
\begin{lem}[Block Matrix Inverse \cite{lu2002inverses}]
Consider the block matrix $\bar{R} \triangleq \left[ {\begin{array}{*{20}{c}}
{\bar A}&{\bar B}\\
{\bar C}&{\bar D}
\end{array}} \right]$. Assume $\bar D$ is nonsingular; then the matrix $\bar{R}$ is invertible if and only if the Schur complement $\bar{A} - \bar{B}\bar{D}^{-1}\bar{C}$  of $\bar{D}$ is invertible, and
\vspace{-0.1cm}
\begin{align}
&{{\bar R}^{ - 1}} \!=\! \left[\!\!\! \begin{array}{l}
{( {\bar A \!-\! \bar B{{\bar D}^{ - 1}}\bar C})^{ - 1}}\\
 - {{\bar D}^{ - 1}}\bar C{( {\bar A \!-\! \bar B{{\bar D}^{ - 1}}\bar C})^{ - 1}}
\end{array} \right. \nonumber\\
&\hspace{2.9cm} \left. \begin{array}{l}
 - {( {\bar A \!-\! \bar B{{\bar D}^{ - 1}}\bar C} )^{ - 1}}\bar B{{\bar D}^{ - 1}}\\
{{\bar D}^{ - 1}} \!+\! {{\bar D}^{ - 1}}\bar C{( {\bar A \!-\! \bar B{{\bar D}^{ - 1}}\bar C})^{ - 1}}\bar B{{\bar D}^{ - 1}}
\end{array} \!\!\!\right]\!\!. \nonumber
\end{align}
\label{tel}
\end{lem}
\vspace{-0.5cm}
\begin{lem}\cite{adamczak2015note}
Let $\mathbf{f}$ be a mean zero random vector in $\mathbb{R}^{n}$, whose covariance matrix is denoted by $\mathrm{Cov}( \mathbf{f} )$. If $\mathbf{f}$ has the convex concentration property with constant $\kappa$, then for any $A \in \mathbb{R}^{n \times n}$ and every $t > 0$, we have
\vspace{-0.2cm}
\begin{align}
\mathbf{P}\!\!\left[ |{{\mathbf{f}^\top}A\mathbf{f} - \mathbf{E}[ {{\mathbf{f}^\top}A\mathbf{f}}]| \ge t}\right] &\le 2{e^{\left( { - \frac{1}{{\mathfrak{c}{\kappa^2}}}\min \left\{ {\frac{{{t^2}}}{{\left\| A \right\|_{\mathrm{F}}^2\mathrm{Cov}( \mathbf{f} )}},\frac{t}{{\left\| A \right\|}}}\right\}}\right)}}\! \nonumber
\end{align}
for some universal constant $\mathfrak{c}$.  \label{newr}
\end{lem}
\vspace{-0.0cm}
\begin{lem}~\cite{vershynin2018high}
Let $W$ be an $d \times d$ a symmetric random matrix. Furthermore, let $\mathcal{N}$ be an $\varepsilon$-net of $\mathcal{S}^{d-1}$ with minimal cardinality. Then for all $\rho > 0$, we have
\begin{align}
&\mathbf{P}\left[|| W || > \rho \right] \nonumber\\
&\le {( {\frac{2}{\varepsilon } + 1})^n}\mathop {\max }\limits_{\mathbf{u} \in \mathcal{N}} \mathbf{P}\left[ {|| W\mathbf{u} ||_{2} > ( {1 - \varepsilon })\rho }\right], ~\varepsilon \in [0, 1).\label{aalem1}\\
&\mathbf{P}\left[|| W || > \rho \right] \nonumber\\
&\le {( {\frac{2}{\varepsilon } \!+\! 1})^n}\mathop {\max }\limits_{\mathbf{u} \in \mathcal{N}} \mathbf{P}\left[ {\left| {{\mathbf{u}^\top}W\mathbf{u}} \right| \!>\! ( {1 \!-\! 2\varepsilon })\rho }\right]\!, ~\varepsilon \in [0, \frac{1}{2}). \label{aalem2}
\end{align}\label{cko}
\end{lem}
\vspace{-0.2cm}
\begin{lem}~\cite{abbasi2011improved}
Let $\{\mathcal{F}_{t}\}_{t \geq 1}$ be a filtration. Let $\{\eta_{t}\}_{t \geq 1}$ be a stochastic process adapted to $\{\mathcal{F}_{t}\}_{t \geq 1}$ and taking values in $\mathbb{R}$. Let $\{x_{t}\}_{t \geq 1}$ be a predictable stochastic process with respect to $\{\mathcal{F}_{t}\}_{t \geq 1}$, taking values in $\mathbb{R}^{d}$. Furthermore, assume that $\eta_{t}$ is $\mathcal{F}_{k}$-measurable and conditionally $\gamma$-sub-Gaussian for some $\gamma > 0$. Let $S > 0$, $\eta^{\top} = [\eta_{2},\eta_{3}, \ldots,\eta_{t+1}]$, and $X^{\top} = [x_{1},x_{2}, \ldots,x_{t}]$. The following
\begin{align}
\left\| \!{{{\left( {{X^ \top }X \!+\! S} \right)}^{ - 0.5}}{X^ \top }\eta } \right\|_2^2 \!\le\! 2{\gamma ^2}\!\ln \!\!\left(\!\! {\frac{{{{\left( {\det\!\! \left(\! {\left( {{X^ \top }X \!+\! S} \right)\!{S^{ - 1}}} \right)} \!\right)}^{0.5}}}}{\delta }} \!\right) \nonumber
\end{align}
holds with the probability of at least $1-\delta$. \label{ckpq}
\end{lem}

\vspace{-0.2cm}
\section*{Appendix B: Proof of Lemma \ref{ko}}
\subsubsection*{\underline{Proof of \eqref{reslem2a}}}We let  ${\bf{w}} \in \ker(P_{(k,p)})$. From \eqref{defmatbdis}, we have
\begin{align}
{{\bf{w}}^\top}P_{(k,p)}{\bf{w}} = \sum\limits_{m = k}^{p - 1} {\sum\limits_{q = \mathbb{T}_m(1)}^{\mathbb{T}_m(\left| {{\mathbb{T}_m}} \right|)} {{\bf{w}}^\top}{\mathbf{r}_m^q{{\left( {\mathbf{r}_m^q} \right)^\top}}{\bf{w}}} } \!=\! 0, \label{hhc}
\end{align}
which means
\begin{align}
( {\mathbf{r}_m^q})^\top{\bf{w}} = 0, ~~~\forall m \in \{k, \ldots, p-1\}, \forall q \in \mathbb{T}_{m},\nonumber
\end{align}
which, in conjunction with \eqref{ddvs}, indicates that
\begin{align}
{\bf{w}} \in \ker (\breve{X}^{\top}_{(k,p)}), ~~~\ker(P_{(k,p)}) = \ker(\breve{X}^{\top}_{(k,p)}). \label{vpk1}
\end{align}

With the consideration of \eqref{deffkh1}, if $\left| {{\mathbb{T}_k}} \right| = p-k$, we have $\mathbb{T}_{k} = \left\{ {k+1,~k+2,~\ldots,~p-1, ~p} \right\}$,
such that each vector of the matrix $\breve{X}_{m}$, defined in \eqref{sufk1}, is a linear combination of vectors of  $\breve{X}_{k}$ for $k+1 \leq m \leq p-1$. Thus, considering \eqref{sufk2}, we have $\ker (\breve{X}^{\top}_{(k,p)}) = \ker (\breve{X}^{\top}_{k})$, which together with \eqref{vpk1} lead to \eqref{reslem2a}.

\subsubsection*{\underline{Proof of \eqref{reslem2b}}}We let  ${\bf{w}} \in \ker(\widehat{X}^{\top}_{(k,p)}\widehat{X}_{(k,p)})$, which means $\mathbf{w}^{\top}\widehat{X}^{\top}_{(k,p)}\widehat{X}_{(k,p)}\mathbf{w} = 0$, which further equates to $\widehat{X}_{(k,p)}\mathbf{w} = \mathbf{0}$. We thus directly obtain \eqref{reslem2b}.

\vspace{-0.3cm}
\section*{Appendix C: Proof of Lemma \ref{cr1}}
For the notation $\mathbf{y}( k ) \triangleq \left[ \begin{array}{l}
\!\!\!{\bf{r}}(k)\!\!\!\!\\
\!\!\!1\!\!\!\!
\end{array} \right]$, we define
\begin{align}
\mathbf{y}^{q}_{m} \triangleq \mathbf{y}( m ) - \mathbf{y}( q ) = \left[ \begin{array}{l}
\!\!\!{\bf{r}}^{q}_{m}\!\!\!\!\\
\!\!\!0\!\!\!\!
\end{array} \right] = \left[ \begin{array}{l}
\!\!\!{\bf{r}}(m) \!-\! {\bf{r}}(q)\!\!\!\!\\
0
\end{array} \right]. \label{dejn1}
\end{align}

If $\text{rank}(\breve{X}_{(k,p)}) < n$, then \eqref{ddvs} implies that there exist scalars $\alpha^{q}_{m}$ $\left(\forall m \in \{k, \ldots, p-1\}, \forall q \in \mathbb{T}_{m} ~\text{with}~ \mathbb{T}_{m} \neq \varnothing\right)$, not all zero, such that
$\sum\limits_{m = k}^{p - 1} {\sum\limits_{q = \mathbb{T}_m(1)}^{\mathbb{T}_m(\left| {{\mathbb{T}_m}} \right|)} {\alpha^{q}_{m}\mathbf{r}_m^q} }  = \mathbf{0}$, which leads to conclude from  \eqref{dejn1} that
\begin{align}
\!\sum\limits_{m = k}^{p - 1} {\sum\limits_{q = \mathbb{T}_m(1)}^{\mathbb{T}_m(\left| {{\mathbb{T}_m}} \right|)}\!\!{\alpha^{q}_{m}\mathbf{y}_m^q} }  = \mathbf{0}_{n+1}. \label{dejn2}
\end{align}

We note that $\widehat{X}_{(k,p)}$ given by \eqref{xf2} can be rewritten as $\widehat{X}_{(k,p)} = [\mathbf{y}( k ), \mathbf{y}( k+1 ), \ldots, \mathbf{y}( p-1 )]^{\top}$. We also note that $\widehat{X}_{(k,p)} \in \mathbb{R}^{(p-k) \times (n+1)}$, and \eqref{dejn2} implies that its rows are linearly dependent. We thus have $\widehat{X}_{(k,p)} < n +1$. In light of \eqref{vpk1} and \eqref{reslem2b}, if  $\text{rank}(P_{(k,p)}) < n$, we have $\text{rank}(\breve{X}_{(k,p)}) < n$, which leads to $\text{rank}(\widehat{X}^{\top}_{(k,p)}\widehat{X}_{(k,p)}) < n + 1$. Finally, by Definition \ref{dfaa} we conclude that if the proposed model inference is not feasible, the naive one is not feasible as well.

\vspace{-0.3cm}
\section*{Appendix D: Proof of Theorem \ref{fthmp}}
\subsubsection*{\underline{Proof of Statement 1)}} Lemma \ref{cr1} implies that to prove statement 1) we only need to prove that if the proposed inference solution  \eqref{self} is feasible, the naive inference solution \eqref{les} is also feasible.

It follows from \eqref{xf2} that
\begin{align}
\widehat{X}^{\top}_{(k,p)}\widehat{X}_{(k,p)} = \left[ {\begin{array}{*{20}{c}}
{\bar A}&{\bar B}\\
{\bar C}&{\bar D}
\end{array}} \right],\label{xfla1}
\end{align}
where
\vspace{-0.1cm}
\begin{align}
\!\!\!\bar{A} \!=\! \sum\limits_{m = k}^{p-1}\! {\mathbf{r}(m)} {\mathbf{r}^\top}(m),~\bar{B} \!=\! \sum\limits_{m = k}^{p-1} \!{\mathbf{r}(m)}  \!=\! {{\bar{C}}^\top},~\bar{D} \!=\! p \!-\! k, \label{xfla2}
\end{align}
from which we obtain that
\vspace{-0.1cm}
\begin{align}
&\bar{A} - \bar{B}{{\bar{D}}^{ - 1}}\bar{C} \nonumber\\
& = \sum\limits_{m = k}^{p-1} {\mathbf{r}(m)} {\mathbf{r}^\top}(m) - \frac{1}{{p-k}}\sum\limits_{m = k}^{p - 1} {\mathbf{r}(m)} \sum\limits_{q = k}^{p - 1} {{\mathbf{r}^\top}(q)} \nonumber\\
& = \frac{1}{{p-k}}\sum\limits_{m = k}^{p - 1}{\sum\limits_{m < q = k + 1}^{p}{(\mathbf{r}(m) \!-\! \mathbf{r}(q)){{(\mathbf{r}(m) \!-\! \mathbf{r}(q))^\top}}} }.\label{pxfla1}
\end{align}

We note that any other vector of nonzero matrix $\breve{X}_{m}$ given in \eqref{sufk1} is a linear combination of the vectors of $\breve{X}_{k}$ with $\left| {{\mathbb{T}_k}} \right| = p-k$, for any $p-1 \geq m \geq k+1$. Thus, we have
\vspace{-0.1cm}
\begin{align}
&\ker  \!\left({\sum\limits_{m < q = k + 1}^{p}{(\mathbf{r}(m) - \mathbf{r}(q)){{(\mathbf{r}(m) - \mathbf{r}(q))^\top}}}}\right) \nonumber\\
&=  \ker  \!\left({\sum\limits_{m < q = k + 1}^{p} \!\!\!\!{\mathbf{r}^{q}_{m}{{(\mathbf{r}^{q}_{m})^\top}}}}\right) = \ker(\breve{X}^{\top}_{k}), ~~\text{if}~\left| {{\mathbb{T}_k}} \right| = p-k, \nonumber
\end{align}
which in light of \eqref{reslem2a} indicates that
\vspace{-0.1cm}
\begin{align}
\ker \left({\sum\limits_{m < q = k + 1}^{p}{\!(\mathbf{r}(m) \!-\! \mathbf{r}(q)){{(\mathbf{r}(m) \!-\! \mathbf{r}(q))^\top}}}}\right) = \ker(P_{(k,p)}), \nonumber
\end{align}
which together with \eqref{pxfla1} further imply that if $P_{(k,p)}$ is invertible, $\bar{A} - \bar{B}{{\bar{D}}^{ - 1}}\bar{C}$ is also invertible. With the consideration of \eqref{xfla1} and Lemma \ref{tel} in Appendix A, we conclude that if \emph{the proposed inference solution \eqref{self} is feasible (i.e., ${P}_{(k,p)}$ is invertible), the naive inference solution \eqref{les} (i.e., ${{\widehat{X}^\top_{(k,p)}}\widehat{X}_{(k,p)}}$ is invertible) is also feasible}. We therefore arrive at the conclusion that the same observation data has identical influence on the feasibility of the two inferences.

\subsubsection*{\underline{Proof of Statement 2)}} Following the method of forming matrix ${P}_{(k,p)}$ in \eqref{npp}, matrix \eqref{defmatbdis1} accordingly updates as
\vspace{-0.1cm}
\begin{align}
{Q}_{(k,p)}  = \sum\limits_{m = k}^{p - 1}{\sum\limits_{m < q = k + 1}^{p}{\mathbf{r}^{{q+1}}_{m+1}{{(\mathbf{r}^{q}_{m})^\top}}} }.\label{npp1}
\end{align}
In addition, we obtain from \eqref{xfla2} that
\vspace{-0.1cm}
\begin{align}
{{\bar D}^{ - 1}}\bar C = \frac{1}{{p-k}}\sum\limits_{m = k}^{p-1} {\mathbf{r}^{\top}( m )}. \label{pxfla3}
\end{align}
It follows from \eqref{xf1} and \eqref{xf2} that
\vspace{-0.1cm}
\begin{align}
\widehat{Y}_{(k,p)}^\top {\widehat{X}_{(k,p)}} = \left[ {\sum\limits_{m = k}^{p-1} {\mathbf{r}(m \!+\! 1){\mathbf{r}^\top}(m)} ,\sum\limits_{m = k}^{p-1} {\mathbf{r}(m \!+\! 1)} }\right].\label{pxfla2}
\end{align}

Since $\widehat{A} = \left[ A,\mathbf{a}\right]$, the structure of inference solution is $\widehat{A}_{\emph{\emph{ls}}} = [A_{\text{ls}},\mathbf{a}_{\text{ls}}]$. Moreover, \eqref{pxfla1} implies that
$\bar{A} - \bar{B}{{\bar{D}}^{ - 1}}\bar{C}$ is invertible if ${P}_{(k,p)}$ is invertible. Thus, in light of Lemma \ref{tel} in Appendix A, with the consideration of \eqref{pxfla1}--\eqref{pxfla2} and the computation \eqref{self}, we have
\vspace{-0.1cm}
\begin{align}
A_{\text{ls}} &= \sum\limits_{m = {k}}^{{p}-1} {\mathbf{r}(m + 1){\mathbf{r}^\top}(m)}(\bar{A} - \bar{B}{{\bar{D}}^{ - 1}}\bar{C})^{-1} \nonumber\\
&\hspace{2.4cm}-\sum\limits_{m = {k}}^{{p}-1} {\mathbf{r}(m \!+\! 1)} {{\bar{D}}^{ - 1}}\bar{C}{( {\bar{A} \!-\! \bar{B}{{\bar{D}}^{ - 1}}\bar{C}})^{ - 1}} \nonumber\\
&=( {p - k} )\sum\limits_{m = k}^{p-1} {\mathbf{r}(m + 1){\mathbf{r}^\top}(m)} P_{(k,p)}^{ - 1} \nonumber\\
&\hspace{2.5cm} - \sum\limits_{m = k}^{p-1} {\mathbf{r}(m + 1)} \sum\limits_{q = m}^{p-1} {{\mathbf{r}^\top}( q )} P_{(k,p)}^{ - 1} \label{aaqq1}\\
& = Q_{(k,p)}P_{(k,p)}^{ - 1} = A_{\text{if}}, \label{aaqq2}
\end{align}
where $P_{(k,p)}$ and $Q_{(k,p)}$ are given by \eqref{npp} and \eqref{npp1}, respectively. The relations  \eqref{aaqq1} and \eqref{aaqq2} indicate that if $\text{rank}({{P}_{(k,p)}}) = n$, the naive and proposed inference procedures generate the same solution of system matrix. Meanwhile, we obtain that
\vspace{-0.15cm}
\begin{align}
\mathbf{a}_{\text{ls}} & = - \sum\limits_{m  = {k}}^{{p} - 1} {\mathbf{r}( {m + 1}){\mathbf{r}^\top}}(m){( {{\bar{A}} - {\bar{B}}{{{\bar{D}}}^{ - 1}}{\bar{C}}})^{ - 1}}{\bar{B}}{{{\bar{D}}}^{ - 1}} \nonumber\\
&\hspace{0.45cm}+ \sum\limits_{m = \overline{k}}^{{p} - 1}\!\! {\mathbf{r}(m + 1)} ( {{{{\bar{D}}}^{ - 1}} \!+\! {{{\bar{D}}}^{ - 1}}{C}{{( {{\bar{A}} \!-\! {\bar{B}}{{{\bar{D}}}^{ - 1}}{\bar{C}}} )}^{ - 1}}{\bar{B}}{{{\bar{D}}}^{ - 1}}} ) \nonumber \\
&= \frac{1}{{{p} - {k}}}\sum\limits_{m = k}^{p - 1} {\left( {\mathbf{r}( {m + 1}) - {Q_{( {k,p})}}P_{( {{k},{p}})}^{ - 1}\mathbf{r}( m)} \right)}, \label{asdq1}\\
&= \frac{1}{{{p} - {k}}}\sum\limits_{m = k}^{p - 1} {\left( {\mathbf{r}( {m + 1}) - A_{\text{if}}\mathbf{r}( m)} \right)} = \mathbf{a}_{\text{if}}, \label{asdq2}
\end{align}
where \eqref{asdq2} from its previous step \eqref{asdq1} is obtained via considering \eqref{aaqq2} and \eqref{selfaa}.

\vspace{-0.00cm}
\section*{Appendix E: Notations of Covariance Matrix }
\vspace{-0.0cm}
\subsubsection*{A. Component $\mathcal{C}_{ww}$} We denote
\vspace{-0.05cm}
\begin{align}
{\mathcal{C}_{ww}} \!\triangleq\! \left[\!\!\! {\begin{array}{*{20}{c}}
{{\mathcal{C}^{w}_{\left( {k,k} \right)}}}\!\!&\!\!{{\mathcal{C}^{w}_{\left( {k,k + 1} \right)}}}\!\!&\!\!{{\mathcal{C}^{w}_{\left( {k,k + 2} \right)}}}\!\!& \!\!\ldots \!\!&\!\!{{\mathcal{C}^{w}_{\left( {k,p - 1} \right)}}}\\
*&{{\mathcal{C}^{w}_{\left( {k+1,k + 1} \right)}}}\!\!&\!\!{{\mathcal{C}^{w}_{\left( {k + 1,k + 2} \right)}}}\!\!&\!\! \ldots \!\!&\!\!{{\mathcal{C}^{w}_{\left( {k + 1,p - 1} \right)}}}\\
*\!\!&\!\!*\!\!&\!\!{{\mathcal{C}_{\left( {k+2,k + 2} \right)}^{w}}}\!\!&\!\! \ldots \!\!&\!\!{{\mathcal{C}^{w}_{\left( {k + 2,p - 1} \right)}}}\\
 \vdots \!\!& \!\!\vdots \!\!&\!\! \vdots \!\!&\!\! \vdots \!\!&\!\! \vdots \\
*\!\!&\!\!*\!\!&\!\!*\!\!&\!\! \ldots \!\!&\!\!{{\mathcal{C}_{\left( {p-1,p-1} \right)}^{w}}}
\end{array}} \!\!\!\right], \nonumber
\end{align}
where we define:
\vspace{-0.05cm}
\begin{align}
&\mathcal{C}_{\left( {k,m,i,j} \right)}^w \triangleq  \begin{cases}
\sigma_{\mathrm{o}}^2{\mathbf{I}_n}, &\text{if}~k = m~\text{or}~k + i = m + j\\
-\sigma_{\mathrm{o}}^2{\mathbf{I}_n}, &\text{if}~k = m+j~\text{or}~k + i = m\\
{\mathbf{O}_{n \times n}}, &\text{otherwise}\\
\end{cases} \nonumber\\
&\mathcal{C}_{\left( {k,m} \right)} \!\triangleq\! \left[\!\!\! {\begin{array}{*{20}{c}}
{\mathcal{C}_{\left( {k,m,{\mathbb{T}_k}\left( 1 \right),{\mathbb{T}_m}\left( 1 \right)} \right)}^w} \!\!&\!\! \ldots \!\!&\!\! {\mathcal{C}_{\left( {k,m,{\mathbb{T}_k}\left( 1 \right),{\mathbb{T}_m}\left( {\left| {{\mathbb{T}_m}} \right|} \right)} \right)}^w}\\
 \vdots \!\!&\!\! \vdots \!\!&\!\! \vdots \\
{\mathcal{C}_{\left( {k,m,{\mathbb{T}_k}\left( {\left| {{\mathbb{T}_k}} \right|} \right),{\mathbb{T}_m}\left( 1 \right)} \right)}^w} \!\!&\!\! \ldots \!\!&\!\! {\mathcal{C}_{\left( {k,m,{\mathbb{T}_k}\left( {\left| {{\mathbb{T}_k}} \right|} \right),{\mathbb{T}_m}\left( {\left| {{\mathbb{T}_m}} \right|} \right)} \right)}^w}
\end{array}} \!\!\!\right]\!\!. \nonumber
\end{align}

\subsubsection*{B. Component $\mathcal{C}_{\breve{f}\breve{f}}$} We denote
\vspace{-0.05cm}
\begin{align}
{\mathcal{C}_{\breve{f}\breve{f}}} \triangleq \left[\!\!\! {\begin{array}{*{20}{c}}
{{\mathcal{C}^{\breve{f}}_{\left( {k,k} \right)}}}\!\!&\!\!{{\mathcal{C}^{\breve{f}}_{\left( {k,k + 1} \right)}}}\!\!&\!\!{{\mathcal{C}^{\breve{f}}_{\left( {k,k + 2} \right)}}}\!\!& \!\!\ldots \!\!&\!\!{{\mathcal{C}^{\breve{f}}_{\left( {k,p - 1} \right)}}}\\
*&{{\mathcal{C}^{\breve{f}}_{\left( {k+1,k + 1} \right)}}}\!\!&\!\!{{\mathcal{C}^{\breve{f}}_{\left( {k + 1,k + 2} \right)}}}\!\!&\!\! \ldots \!\!&\!\!{{\mathcal{C}^{\breve{f}}_{\left( {k + 1,p - 1} \right)}}}\\
*\!\!&\!\!*\!\!&\!\!{{\mathcal{C}_{\left( {k+2,k + 2} \right)}^{\breve{f}}}}\!\!&\!\! \ldots \!\!&\!\!{{\mathcal{C}^{\breve{f}}_{\left( {k + 2,p - 1} \right)}}}\\
 \vdots \!\!& \!\!\vdots \!\!&\!\! \vdots \!\!&\!\! \vdots \!\!&\!\! \vdots \\
*\!\!&\!\!*\!\!&\!\!*\!\!&\!\! \ldots \!\!&\!\!{{\mathcal{C}_{\left( {p-1,p-1} \right)}^{\breve{f}}}}
\end{array}} \!\!\!\right], \nonumber
\end{align}
where we define:
\begin{align}
&\mathcal{C}^{\breve{f}}_{\left( {k,m,i,j} \right)} \triangleq  \begin{cases}
\sigma_{\mathrm{p}}^2{\mathbf{I}_n}, &\text{if}~k = m~\text{or}~k - i = m - j\\
-\sigma_{\mathrm{p}}^2{\mathbf{I}_n}, &\text{if}~k = m-j~\text{or}~k - i = m\\
{\mathbf{O}_{n \times n}}, &\text{otherwise}\\
\end{cases} \nonumber\\
&\mathcal{C}_{\left( {k,m} \right)}^{\breve{f}} \triangleq \left[\!\!\! {\begin{array}{*{20}{c}}
{\mathcal{C}_{\left( {k,m,1,1} \right)}^{\breve{f}}}\!\!&\!\!{\mathcal{C}_{\left( {k,m,1,2} \right)}^{\breve{f}}} \!\!&\!\! \ldots \!\!&\!\! {\mathcal{C}_{\left( {k,m,1,k - 1} \right)}^{\breve{f}}}\\
{\mathcal{C}_{\left( {k,m,2,1} \right)}^{\breve{f}}}\!\!&\!\!{\mathcal{C}_{\left( {k,m,2,2} \right)}^{\breve{f}}} \!\!&\!\! \ldots \!\!&\!\! {\mathcal{C}_{\left( {k,m,2,k - 1} \right)}^{\breve{f}}}\\
 \vdots \!\!&\!\! \vdots \!\!&\!\! \vdots \!\!&\!\! \vdots \\
{\mathcal{C}_{\left( {k,m,k - 1,1} \right)}^{\breve{f}}}\!\!&\!\!{\mathcal{C}_{\left( {k,m,k - 1,2} \right)}^{\breve{f}}}\!\!&\!\! \ldots \!\!&\!\!{\mathcal{C}_{\left( {k,m,k - 1,k - 1} \right)}^{\breve{f}}}
\end{array}} \!\!\!\right]\!. \nonumber
\end{align}

\subsubsection*{C. Component ${\mathcal{C}_{\widetilde{f} \widetilde {f}}}$} We denote
\begin{align}
{\mathcal{C}_{\widetilde{f} \widetilde {f}}} \triangleq \left[\!\!\! {\begin{array}{*{20}{c}}
{{\mathcal{C}^{\widetilde{f}}_{\left( {k,k} \right)}}}\!\!&\!\!{{\mathcal{C}^{\widetilde{f}}_{\left( {k,k + 1} \right)}}}\!\!&\!\!{{\mathcal{C}^{\widetilde{f}}_{\left( {k,k + 2} \right)}}}\!\!& \!\!\ldots \!\!&\!\!{{\mathcal{C}^{\widetilde{f}}_{\left( {k,p - 1} \right)}}}\\
*&{{\mathcal{C}^{\widetilde{f}}_{\left( {k+1,k + 1} \right)}}}\!\!&\!\!{{\mathcal{C}^{\widetilde{f}}_{\left( {k + 1,k + 2} \right)}}}\!\!&\!\! \ldots \!\!&\!\!{{\mathcal{C}^{\widetilde{f}}_{\left( {k + 1,p - 1} \right)}}}\\
*\!\!&\!\!*\!\!&\!\!{{\mathcal{C}_{\left( {k+2,k + 2} \right)}^{\widetilde{f}}}}\!\!&\!\! \ldots \!\!&\!\!{{\mathcal{C}^{\widetilde{f}}_{\left( {k + 2,p - 1} \right)}}}\\
 \vdots \!\!& \!\!\vdots \!\!&\!\! \vdots \!\!&\!\! \vdots \!\!&\!\! \vdots \\
*\!\!&\!\!*\!\!&\!\!*\!\!&\!\! \ldots \!\!&\!\!{{\mathcal{C}_{\left( {p-1,p-1} \right)}^{\widetilde{f}}}}
\end{array}} \!\!\!\right], \nonumber
\end{align}
where we define:
\begin{align}
&\mathcal{C}_{\left( {m,i,j} \right)}^{\widetilde f} \triangleq \begin{cases}
(\sigma_{\mathrm{p}}^2+\sigma_{\mathrm{a}}^2){\mathbf{I}_n}, &\text{if}~m - i = m - j\\
{\mathbf{O}_{n \times n}}, &\text{otherwise}\\
\end{cases} \nonumber\\
&\mathcal{C}_{\left( {k,m} \right)}^{\widetilde f} \!\triangleq\! \left[\!\!\! {\begin{array}{*{20}{c}}
{\mathcal{C}_{\left( {m,k,k} \right)}^{\widetilde f}}\!&\!{\mathcal{C}_{\left( {m,k,k + 1} \right)}^{\widetilde f}}\!&\! \ldots \!&\!{\mathcal{C}_{\left( {m,k,m - 1} \right)}^{\widetilde f}}\\
{\mathcal{C}_{\left( {m,k + 1,k} \right)}^{\widetilde f}}\!&\!{\mathcal{C}_{\left( {m,k + 1,k + 1} \right)}^{\widetilde f}}\!&\! \ldots \!&\!{\mathcal{C}_{\left( {m,k + 1,m - 1} \right)}^{\widetilde f}}\\
 \vdots \!&\! \vdots \!&\! \vdots \!&\! \vdots \\
{\mathcal{C}_{\left( {m,m - 1,k} \right)}^{\widetilde f}}\!&\!{\mathcal{C}_{\left( {m,m - 1,k + 1} \right)}^{\widetilde f}}\!&\! \ldots \!&\!{\mathcal{C}_{\left( {m,m - 1,m - 1} \right)}^{\widetilde f}}
\end{array}} \!\!\!\right]\!\!. \nonumber
\end{align}

\subsubsection*{D. Component ${\mathcal{C}_{\breve{f}\widetilde{f}}}$} We denote
\begin{align}
{\mathcal{C}_{\breve{f}\widetilde{f}}} = \left[ {\begin{array}{*{20}{c}}
{{\breve{\mathcal{C}}_{\left( {k,k} \right)}}}&{{\breve{\mathcal{C}}_{\left( {k,k + 1} \right)}}}& \ldots &{{\breve{\mathcal{C}}_{\left( {k,p - 1} \right)}}}\\
{{\breve{\mathcal{C}}_{\left( {k + 1,k} \right)}}}&{{\breve{\mathcal{C}}_{\left( {k + 1,k + 1} \right)}}}& \ldots &{{\breve{\mathcal{C}}_{\left( {k + 1,p - 1} \right)}}}\\
 \vdots & \vdots & \vdots & \vdots \\
{{\breve{\mathcal{C}}_{\left( {p - 1,k} \right)}}}&{{\breve{\mathcal{C}}_{\left( {p - 1,k + 1} \right)}}}& \ldots &{{\breve{\mathcal{C}}_{\left( {p - 1,p - 1} \right)}}}
\end{array}} \right], \nonumber
\end{align}
where we define:
\begin{align}
&{\breve{\mathcal{C}}_{\left( {k,m,i,j,s,t} \right)}} \triangleq \begin{cases}
- \sigma_{\mathrm{p}}^2{\mathbf{I}_n}, \!\!&{\mathbb{T}_k}\left( i \right) \!-\! 1 \!-\! s \!=\! {\mathbb{T}_m}\left( j \right) \!-\! m \!-\! t\\
  \sigma_{\mathrm{p}}^2{\mathbf{I}_n}, \!\!&k \!-\! 1 \!-\! s \!=\! {\mathbb{T}_m}\left( j \right) \!-\! m \!-\! t\\
  \mathbf{O}_{n \times n},  \!\!&\text{otherwise}
\end{cases} \nonumber\\
&{\breve{\mathcal{C}}_{\left( {k,m,i,j} \right)}} \!\triangleq\! \left[ {\begin{array}{*{20}{c}}
{{\breve{\mathcal{C}}_{\left( {k,m,i,j,0,0} \right)}}} & \ldots & {{\breve{\mathcal{C}}_{\left( {k,m,i,j,0,{\mathbb{T}_m}\left( j \right) - j} \right)}}}\\
 \vdots & \vdots & \vdots \\
{{\breve{\mathcal{C}}_{\left( {k,m,i,j,k - 2,0} \right)}}} & \ldots & {{\breve{\mathcal{C}}_{\left( {k,m,i,j,k - 2,{\mathbb{T}_m}\left( j \right) - j} \right)}}}
\end{array}} \right]\!\!,\nonumber\\
&{\breve{\mathcal{C}}_{\left( {k,m} \right)}} \!\triangleq\! \left[\!\! {\begin{array}{*{20}{c}}
{{\breve{\mathcal{C}}_{\left( {k,m,{\mathbb{T}_k}\left( 1 \right),{\mathbb{T}_m}\left( 1 \right)} \right)}}} \!\!&\!\! \ldots \!\!&\!\! {{\breve{\mathcal{C}}_{\left( {k,m,{\mathbb{T}_k}\left( 1 \right),{\mathbb{T}_m}\left( {\left| {{\mathbb{T}_m}} \right|} \right)} \right)}}}\\
 \vdots \!\!&\!\! \vdots \!\!&\!\! \vdots \\
{{\breve{\mathcal{C}}_{\left( {k,m,{\mathbb{T}_k}\left( {\left| {{\mathbb{T}_k}} \right|} \right),{\mathbb{T}_m}\left( 1 \right)} \right)}}} \!\!&\!\! \ldots \!\!&\!\! {{\breve{\mathcal{C}}_{\left( {k,m,{\mathbb{T}_k}\left( {\left| {{\mathbb{T}_k}} \right|} \right),{\mathbb{T}_m}\left( {\left| {{\mathbb{T}_m}} \right|} \right)} \right)}}}
\end{array}} \!\!\right]\!\!. \nonumber
\end{align}

\section*{Appendix F: Proof of Proposition \ref{ptopk1a}}
Let us denote
\begin{align}
\mathbf{q} &\triangleq \frac{1}{p - k} {\sum\limits_{m = k}^{p - 1} \!{( {\mathbf{f}(m) + \mathbf{w}(m + 1) - A\mathbf{w}(m)})} },\label{kmcf1}\\
\widehat{\mu} &\triangleq \mu {\mathbf{1}_n} - A\mu {\mathbf{1}_n}, ~~~~\overline{\mathbf{r}} \triangleq \frac{1}{p-k}\sum\limits_{m = k}^{p - 1}{{\mathbf{r}(m)}}.\label{kmcf2kk}
\end{align}

\subsubsection*{\underline{Proof of \eqref{hhhccc3}}} Considering $\overline{\mathbf{r}}$ in \eqref{kmcf2kk}, we obtain from \eqref{unda}:
\begin{align}
&\overline{\mathbf{r}} - \mu {\mathbf{1}_n} = \overline{\mathbf{w}} + \frac{1}{{p - k}}\sum\limits_{m = k}^{p - 1} {{A^{m - 1}}x(1)}  + \frac{1}{{p - k}}\sum\limits_{m = k}^{p - 1} {\sum\limits_{i = 0}^{m - 2} {{A^i}} \mathbf{a}} \nonumber\\
&\hspace{2.0cm} + \frac{1}{{p - k}}\sum\limits_{m = 1}^{p - 2} {\sum\limits_{i = k - 1 - m}^{p - 2 - m} {\mathfrak{u}(i){A^i}\mathbf{f}(m)}}, \label{pccv}
\end{align}
where $\overline{\mathbf{w}} \triangleq \frac{1}{p-k}\sum\limits_{m = k}^{p - 1}{{\mathbf{w}(m)}} - \mu {\mathbf{1}_n}$. From \eqref{pccv} and Assumption \ref{asprv}-1) we verify that
\begin{align}
&{\left\| {\overline{\mathbf{r}} - \mu {\mathbf{1}_n}} \right\|^2} - \overline{\mathfrak{h}} = {( {\overline{\mathbf{r}} - \mu {\mathbf{1}_n}})^2}{\mathbf{I}_n}( {\overline{\mathbf{r}} - \mu {\mathbf{1}_n}}) \nonumber\\
 &\hspace{3.0cm}- \mathbf{E}\left[ {{{( {\overline{\mathbf{r}} - \mu {\mathbf{1}_n}})^2}}{\mathbf{I}_n}( {\overline{\mathbf{r}} - \mu {\mathbf{1}_n}})} \right], \label{pq1}\\
& \text{Cov}({\overline{\mathbf{r}} - \mu {\mathbf{1}_n}}) = \underline{\mathcal{C}}, \label{pq2}
\end{align}
where $\overline{\mathfrak{h}}$ and $\underline{\mathcal{C}}$ are given in \eqref{kmcf2} and \eqref{kmcf3}, respectively. We note that Assumptions \ref{asprv}-1) and \ref{asprv}-2) imply that ${\overline{\mathbf{r}} - \mu {\mathbf{1}_n}}$ has zero mean and convex concentration property with constant $\kappa > 0$. Let us set $\mathfrak{c} > 0$ and $\rho_{1} > 0$. Applying Lemma \ref{newr}, with consideration of $\left\| {{\mathbf{I}_n}} \right\|_{\mathrm{F}}^2 = n$, $\left\| {{\mathbf{I}_n}} \right\| = 1$, \eqref{pq1} and \eqref{pq2}, we obtain \eqref{hhhccc3}.

\subsubsection*{\underline{Proof of \eqref{hhhccc2}}} Considering Assumption \ref{asprv}-1), the relation $\left\| A \right\|_{\mathrm{F}}^2 = \text{tr}( {{A^\top}A})$ and the defined vectors $\mathbf{q}$ in \eqref{kmcf1} and $\widehat{\mu}$ in \eqref{kmcf2kk}, we have
\begin{align}
{\left\| \mathbf{q} - \widehat{\mu}\right\|^2} - \underline{\mathfrak{h}} &= {\left\|  \mathbf{q} - \widehat{\mu} \right\|^2} - \mathbf{E}[ {{{\left\|  \mathbf{q} - \widehat{\mu} \right\|}^2}}] \label{pmcca1}\\
& = {( \mathbf{q} \!-\! \widehat{\mu})^\top}{\mathbf{I}_n}( \mathbf{q} \!-\! \widehat{\mu}) \!-\! \mathbf{E}[ {{{( \mathbf{q} \!-\! \widehat{\mu})^\top}}{\mathbf{I}_n}( \mathbf{q} \!-\! \widehat{\mu})}], \nonumber\\
&\hspace{-2.0cm}\text{Cov}(\mathbf{q} - \widehat{\mu}) = \overline{\mathcal{C}}, \label{pmcca2}
\end{align}
where $\overline{\mathcal{C}}$ and $\underline{\mathfrak{h}}$ are given in \eqref{kmcf1ab} and \eqref{kmcfa2}, respectively. We note that Assumptions \ref{asprv}-1) and \ref{asprv}-2) implies that $\mathbf{q} - \widehat{\mu}$ has zero mean and convex concentration property with constant $\kappa > 0$. Let us set $\mathfrak{c} > 0$ and $\rho_{2} > 0$. Applying Lemma \ref{newr}, with consideration of $\left\| {{\mathbf{I}_n}} \right\|_{\mathrm{F}}^2 = n$, $\left\| {{\mathbf{I}_n}} \right\| = 1$, \eqref{pmcca1} and \eqref{pmcca2}, we obtain \eqref{hhhccc2}.

\subsubsection*{\underline{Proof of \eqref{hhhccc1}}}
It follows from \eqref{kop3p} that
\begin{align}
&\mathbf{E}\!\left[\! {M_{(k,p)}^ \top {X_{(k,p)}}X_{(k,p)}^ \top {M_{(k,p)}}} \!\right]  \!=\! M_{(k,p)}^ \top \mathbf{E}\!\left[ {{X_{(k,p)}}X_{(k,p)}^ \top } \right]\!{M_{(k,p)}} \nonumber\\
&= M_{(k,p)}^ \top {\Gamma _{(k,p)}}{M_{(k,p)}} = \mathbf{I}_{n}.  \label{trans2}
\end{align}

We note for a matrix $A \in \mathbb{R}^{n \times n}$, $\left\| A \right\| = \mathop {\sup }\limits_{\mathbf{u} \in {\mathcal{S}^{n - 1}}} \left| {{\mathbf{u}^\top}A\mathbf{u}} \right|$, which, in conjunction with \eqref{trans2}, leads to
\begin{align}
&\left\| {{{( {X_{(k,p)}^ \top {M_{(k,p)}}})^\top}}X_{(k,p)}^ \top {M_{(k,p)}} - \mathbf{I}_{n}} \right\| \nonumber\\
& = \left\| {{{( {X_{(k,p)}^ \top {M_{(k,p)}}})^\top} }X_{(k,p)}^ \top {M_{(k,p)}}} \right. \nonumber\\
&\hspace{3.8cm}\left. { - \mathbf{E}[ {M_{(k,p)}^ \top {X_{(k,p)}}X_{(k,p)}^ \top {M_{(k,p)}}}]} \right\| \nonumber\\
& = \mathop {\sup }\limits_{\mathbf{u} \in {\mathcal{S}^{n - 1}}} \left| {{\mathbf{u}^\top}{{( {X_{(k,p)}^\top {M_{(k,p)}}})^\top}}X_{(k,p)}^ \top {M_{(k,p)}}\mathbf{u}} \right.\nonumber\\
&\hspace{3.1cm}\left. { - {\mathbf{u}^\top}\mathbf{E}[ {M_{(k,p)}^\top {X_{(k,p)}}X_{(k,p)}^ \top {M_{(k,p)}}}]\mathbf{u}} \right|\nonumber\\
& = \mathop {\sup }\limits_{\mathbf{u} \in {\mathcal{S}^{n - 1}}} \left| {\left\| {X_{(k,p)}^ \top {M_{(k,p)}}\mathbf{u}} \right\|_2^2 \!-\! \mathbf{E}\left\| {X_{(k,p)}^ \top {M_{(k,p)}}\mathbf{u}} \right\|_2^2} \right|\!.  \label{gh1ab}
\end{align}

Let us define:
\begin{align}
\mathbf{U} \triangleq \text{diag}\left\{ {\mathbf{u}},~{\mathbf{u}},~\ldots,~{\mathbf{u}} \right\} \in \mathbb{R}^{n\mathfrak{n}_{(k,p)} \times \mathfrak{n}_{(k,p)}},  \label{kmccq}
\end{align}
where $\mathfrak{n}_{(k,p)}$ is given by \eqref{gh9bb}. Observing \eqref{ghbvector} with \eqref{ghb} and \eqref{ghmatrix} with \eqref{gh}, we arrive at
\begin{align}
X_{({k},{p})}^\top {M_{({k},{p})}}\mathbf{u} = \mathbf{U}^{\top}{{\Upsilon}^{\top}_{({k},{p})}}{{\Pi}_{( {{k},{p}})}}{\eta_{({k},{p})}}, \label{trans1}
\end{align}
where ${\Upsilon_{(k,p)}}$, ${\Pi _{( {k,p})}}$ and ${\eta _{(k,p)}}$ are defined by \eqref{gh9}, \eqref{ghmatrix} and \eqref{ghbvector}, respectively. Inserting \eqref{trans1} into \eqref{gh1ab}, we arrive at \begin{align}
&\left\| {{{( {X_{({k},{p})}^ \top {M_{({k},{p})}}})^\top} }X_{({k},{p})}^ \top {M_{({k},{p})}} - \mathbf{I}_{n}} \right\|  \nonumber\\
& = \mathop {\sup }\limits_{\mathbf{u} \in {\mathcal{S}^{n - 1}}} \left| {\left\| \mathbf{U}^{\top}{{\Upsilon}^{\top}_{({k},{p})}}{{\Pi}_{( {{k},{p}})}}{{\eta}_{({k},{p})}} \right\|_2^2} \right. \nonumber\\
&\hspace{3.0cm} \left. { - \mathbf{E}\!\left\| \mathbf{U}^{\top}{{\Upsilon}^{\top}_{({k},{p})}}{{\Pi}_{( {{k},{p}})}}{{\eta}_{({k},{p})}} \right\|_2^2} \right|\!. \label{relat}
\end{align}

Let us define:
\begin{align}
\Delta_{({k},{p})}  \triangleq {{\Pi}^{\top}_{({{k},{p}})}}{{\Upsilon}_{({k},{p})}}\mathbf{U}\mathbf{U}^{\top}{{\Upsilon}^{\top}_{({k},{p})}}{{\Pi}_{( {{k},{p}})}},\label{ghb1ok11}
\end{align}
which follows from $\mathbf{u} \in {\mathcal{S}^{n - 1}}$, \eqref{kmccq}, \eqref{gh9bb} and the well-known inequalities $|| {AB} ||$ $\le$ $|| A|||| B ||$ and $|| {AB} ||_{\mathrm{F}}$ $\le$ $|| A |||| B ||_{\mathrm{F}}$ that
\begin{align}
&||\Delta_{({k},{p})}\! ||_{\mathbf{F}} \leq ||{{\Pi}^{\top}_{({{k},{p}})}}{{\Upsilon}_{({k},{p})}}||^{2}||\mathbf{U}||||\mathbf{U}||_{\mathbf{F}} \nonumber\\
&= ||{{\Pi}^{\top}_{({{k},{p}})}}\!{{\Upsilon}_{({k},{p})}}||^{2}\sqrt{\mathfrak{n}_{(k,p)}},\label{mm1}\\
&||\Delta_{({k},{p})}|| \!\leq\! ||{{\Pi}^{\top}_{({{k},{p}})}}{{\Upsilon}_{({k},{p})}}||^{2}||\mathbf{U}\mathbf{U}^{\top}\!|| \!\leq\!  ||{{\Pi}^{\top}_{({{k},{p}})}}{{\Upsilon}_{({k},{p})}}||^{2}\!, \label{mm2}
\end{align}
where \eqref{mm1} and \eqref{mm2} from their previous steps are obtained via considering $||\mathbf{U}||_{\mathbf{F}} = \sqrt{\mathfrak{n}_{(k,p)}}$ and $||\mathbf{U}|| = 1$. In light of \eqref{ghb1ok11}--\eqref{mm2}, with  $\rho_{3} > 0$, we have
\begin{align}
&\min \!\left\{ {\frac{{{\rho_{3} ^2}}}{{|| \Delta_{({k},{p})} ||_\mathrm{F}^2|| {\mathcal{C}}_{\mathrm{v}} ||}},~\frac{{\rho_{3}}}{{|| \Delta_{({k},{p})} ||}}} \right\} \nonumber\\
&\!\geq\! \min\! \left\{\!\! {\frac{{{\rho_{3}^2}}}{{\mathfrak{n}_{({k},{p})}||{{\Pi}^{\top}_{({{k},{p}})}}\!{{\Upsilon}_{({k},{p})}}||^{4}|| {\mathcal{C}}_{\mathrm{v}} ||}},\frac{{\rho_{3}}}{{|| {{\Pi}^{\top}_{({{k},{p}})}}\!{{\Upsilon}_{({k},{p})}} ||^{2}}}} \!\!\right\}\!. \label{ghb1ok}
\end{align}

Let us set $\mathfrak{c} > 0$. Applying Lemma \ref{newr}, with consideration of \eqref{ghb1ok}, \eqref{ghb1ok11} and \eqref{kop2}, we conclude that
\begin{align}
\left| {\left\| {\Upsilon_{(k,p)}^ \top {\Pi _{(k,p)}}{\eta _{(k,p)}}} \right\|_2^2 - {\bf{E}}\left\| {\Upsilon _{(k,p)}^ \top {\Pi _{(k,p)}}{\eta _{(k,p)}}} \right\|_2^2} \right| > \rho_{3} \nonumber
\end{align}
holds with probability at most
\begin{align}
& 2e^{- \frac{1}{{\mathfrak{c}{\kappa^2}}}\min \left\{ {\frac{{{\rho_{3}^2}}}{{|| \Delta_{({k},{p})} ||_\mathrm{F}^2|| {\mathcal{C}}_{\mathrm{v}} ||}},~\frac{{\rho_{3}}}{{|| \Delta_{({k},{p})} ||}}} \right\}} \nonumber \\
& \leq 2e^{- \frac{1}{{\mathfrak{c}{\kappa^2}}}\min\! \left\{\!\! {\frac{{{\rho_{3} ^2}}}{{\mathfrak{n}_{({k},{p})}||{{\Pi}^{\top}_{({{k},{p}})}}\!{{\Upsilon}_{({k},{p})}}||^{4}|| {\mathcal{C}}_{\mathrm{v}} ||}},~\frac{{\rho_{3}}}{{|| {{\Pi}^{\top}_{({{k},{p}})}}\!{{\Upsilon}_{({k},{p})}} ||^{2}}}} \!\!\right\}}. \nonumber
\end{align}
Then, noting \eqref{relat}, as a consequence, we have that
\begin{align}
&\left\| {{\left( {X_{(k,p)}^ \top {M_{(k,p)}}} \right)^\top}X_{(k,p)}^ \top {M_{(k,p)}} - {\bf{I}_{n}}} \right\| \nonumber\\
& = \mathop {\sup }\limits_{\mathbf{u} \in {\mathcal{S}^{n - 1}}} \left| {{\mathbf{u}^\top}\left( {{{\left( {X_{(k,p)}^ \top {M_{(k,p)}}} \right)}^\top }X_{(k,p)}^ \top {M_{(k,p)}} - {\bf{I}_{n}}} \right)\mathbf{u}} \right| \nonumber\\
&\ge (1 - 2\varepsilon )\rho_{3} \nonumber
\end{align}
holds with probability at most
\begin{align}
2e^{- \frac{1}{{\mathfrak{c}{\kappa^2}}}\min \left\{ {\frac{{{(1 - 2\varepsilon )^2\rho_{3} ^2}}}{{{{{\mathfrak{n}_{({k},{p})}||{{\Pi}^{\top}_{({{k},{p}})}}\!{{\Upsilon}_{({k},{p})}}||^{4}}}}|| {{{\mathcal{C}}_\mathrm{v}}} ||}},~\frac{{(1 - 2\varepsilon)\rho_{3}}}{{{||{\Pi}^{\top}_{({{k},{p}})}{\Upsilon}_{({k},{p})}} ||^{2}}}} \right\}}. \nonumber
\end{align}
Then, applying \eqref{aalem2} in Lemma \ref{cko}, we obtain \eqref{hhhccc1}.

\section*{Appendix G: Proof of Theorem \ref{ptopk1}}
We first derive two inequalities, which will be used to prove Statements 1) and 2). We conclude from \eqref{self} and \eqref{lm3r} that $A - A_{\text{if}} = -R_{(k,p)}{P^{ - 1}_{(k,p)}}$. Noticing \eqref{defmatbdis}, we then have
\begin{align}
&\left\| {A - {A_{{\rm{if}}}}} \right\|  \nonumber\\
& = \left\| {{R_{(k,p)}}P_{(k,p)}^{ - 1}} \right\| \nonumber\\
& = \left\| {{R_{(k,p)}}P_{(k,p)}^{ - 0.5}P_{(k,p)}^{ - 0.5}} \right\| \nonumber\\
& \le \left\| {{R_{(k,p)}}P_{(k,p)}^{ - 0.5}} \right\|\left\| {P_{(k,p)}^{ - 0.5}} \right\| \nonumber\\
& = \left\| {{R_{(k,p)}}{{( {{{X}_{(k,p)}}{X}_{(k,p)}^ \top })^{ - 0.5}}}} \right\|\left\| {{{( {{{X}_{(k,p)}}X_{(k,p)}^ \top })^{ - 0.5}}}} \right\|. \label{ckko1}
\end{align}
Meanwhile, we obtain from \eqref{un} that
\begin{align}
\alpha  \!=\! \frac{1}{{p \!-\! k}}\!\sum\limits_{m = k}^{p - 1} \!\!{( {\mathbf{r}(m \!+\! 1) \!-\! A\mathbf{r}(m) \!-\! \mathbf{f}(m) \!+\! A\mathbf{w}(m) \!-\! \mathbf{w}(m + 1)})}\nonumber
\end{align}
which, in conjunction with \eqref{selfaa}, \eqref{kmcf1} and \eqref{kmcf2kk}, leads to ${\alpha _{\text{if}}} - \alpha = ( {A - A_{\text{if}}})\overline{\mathbf{r}} + \mathbf{q}$, as a consequence, we have
\begin{align}
\left\| {{\alpha _{\text{if}}} - \alpha} \right\| \le \left\| {A - A_{\text{if}}} \right\|\left\| \overline{\mathbf{r}} \right\| + \left\| \mathbf{q} \right\|.\label{ckkoba1}
\end{align}

\subsection{Proof of Statement 1)}
For \eqref{ckko1},  we define two events:
\begin{subequations}
\begin{align}
{\Omega _1} &\triangleq \left\{ {\left\| {{R_{(k,p)}}{{\left( {{X_{(k,p)}}X_{(k,p)}^ \top } \right)}^{ - 0.5}}} \right\|} \right.\nonumber\\
&\hspace{2.5cm}\left. { \cdot \left\| {{{\left( {{X_{(k,p)}}X_{(k,p)}^ \top } \right)}^{ - 0.5}}} \right\| > \phi } \right\},\\
{\Omega _2} &\triangleq \left\{ {\left\| {M_{(k,p)}^ \top {X_{(k,p)}}X_{(k,p)}^ \top {M_{(k,p)}} - \mathbf{I}_{n}} \right\| \le \rho_{3} } \right\},
\end{align}\label{event}
\end{subequations}
\!\!from which, using \eqref{ckko1}, we have
\begin{align}
\mathbf{P}\!\left[ {\left\| {{A_{{\rm{if}}}} - A} \right\| > \phi } \right] \le \mathbf{P}\!\left[ {{\Omega _1}\bigcap {{\Omega _2}} } \right] + \mathbf{P}\!\left[ {\Omega _2^\mathrm{c}} \right]. \label{kp1}
\end{align}
We next derive two upper bounds.

\subsubsection*{\underline{Upper Bound on $\mathbf{P}( {\Omega _2^\mathrm{c}})$}} We let
\begin{align}
\gamma  = \sqrt {2\mathfrak{c}} \kappa,\label{kp1co}
\end{align}
inserting which into \eqref{fg0aa} results in
\begin{align}
&\min \left\{ {\frac{{{(1 - 2\varepsilon )^2\rho_{3}^2}}}{{{{{\mathfrak{n}_{({k},{p})}||{{\Pi}^{\top}_{({{k},{p}})}}{{\Upsilon}_{({k},{p})}}||^{4}}}}|| {{{\mathcal{C}}_\mathrm{v}}} ||}},~\frac{{(1 - 2\varepsilon)\rho_{3}}}{{{||{\Pi}^{\top}_{({{k},{p}})}{\Upsilon}_{({k},{p})}} ||^{2}}}} \right\} \nonumber \\
&\ge \mathfrak{c}{\kappa ^2}\ln \frac{{4 \cdot {( {\frac{2}{\varepsilon } + 1})^n}}}{\delta }, \nonumber
\end{align}
which is equivalent to
\begin{align}
2 \!\cdot\! {( {\frac{2}{\varepsilon } \!+\! 1})^n} \!\cdot\! {e^{\frac{-1}{{\mathfrak{c}{\kappa^2}}}\!\min \left\{\!\! {\frac{{{(1 \!-\! 2\varepsilon )^2\rho_{3} ^2}}}{{{{{\mathfrak{n}_{({k},{p})}||{{\Pi}^{\top}_{({{k},{p}})}}\!{{\Upsilon}_{({k},{p})}}||^{4}}}}|| {{{\mathcal{C}}_\mathrm{v}}} ||}},\frac{{(1 \!-\! 2\varepsilon)\rho_{3}}}{{{||{\Pi}^{\top}_{({{k},{p}})}\!{\Upsilon}_{({k},{p})}} ||^{2}}}} \!\!\right\}}} \!\!\le\! \frac{\delta }{2} \nonumber
\end{align}
which together with \eqref{hhhccc1} in Proposition \ref{ptopk1a} imply that when \eqref{fg0aa} holds, we have
\begin{align}
\mathbf{P}( {\Omega _2^\mathrm{c}} ) \leq \frac{\delta}{2}. \label{kpp1}
\end{align}

\subsubsection*{\underline{Upper Bound  on $\mathbf{P}( {{\Omega _1}\bigcap {{\Omega _2}} } )$}} When ${\Omega _2}$ occurs, we have
\begin{align}
( {1 - \rho_{3} })\mathbf{I}_{n} \le M_{(k,p)}^ \top {X_{(k,p)}}X_{(k,p)}^ \top {M_{(k,p)}} \le ( {1 + \rho_{3} })\mathbf{I}_{n}. \nonumber
\end{align}
As a consequence, we get
\begin{align}
( {1 - \rho_{3} })M^{-2}_{(k,p)} \le {X_{(k,p)}}X_{(k,p)}^\top \le ( {1 + \rho_{3} })M^{-2}_{(k,p)}, \label{fg1}
\end{align}
which implies
\begin{align}
\lambda^{0.5}_{\min}({X_{(k,p)}}X_{(k,p)}^\top) \geq \lambda^{0.5}_{\min}(( {1 - \rho_{3} }){M^{-2}_{(k,p)}}) \triangleq \beta. \label{fg1j0}
\end{align}

We note that
\begin{align}
|| {{{( {{X_{(k,p)}}X_{(k,p)}^ \top })^{0.5}}}} || \!=\! \lambda _{\max }^{0.5}\!( {{X_{(k,p)}}X_{(k,p)}^ \top }) \!\ge\! \lambda _{\min }^{0.5}\!( {{X_{(k,p)}}X_{(k,p)}^\top })\nonumber
\end{align}
by which, we thus obtain from \eqref{fg1j0} that
\begin{align}
\frac{1}{\beta } \ge || {{{( {{X_{(k,p)}}X_{(k,p)}^ \top })^{ - 0.5}}}} ||.\nonumber
\end{align}
As a consequence, we conclude from \eqref{event} that
\begin{align}
\!\!{\Omega _1} \bigcap {\Omega_2} \subseteq\! \left\{\! {||{{R_{(k,p)}}{{( {{X_{(k,p)}}X_{(k,p)}^ \top })^{ - 0.5}}}} || \!>\! \beta \phi } \!\right\} \!\bigcap {\Omega_2}. \label{fg2}
\end{align}

The left-hand inequality of \eqref{fg1} implies
\begin{align}
2{X_{(k,p)}}X_{(k,p)}^ \top  \ge ( {1 - \rho_{3} })M_{(k,p)}^{ - 2} + {X_{(k,p)}}X_{(k,p)}^ \top, \nonumber
\end{align}
which means, with $0 < \rho_{3} < 1$, that
\begin{align}
{( {{X_{(k,p)}}X_{(k,p)}^ \top })^{ - 1}} \le 2{( {( {1 - \rho_{3} })M_{(k,p)}^{ - 2} + {X_{(k,p)}}X_{(k,p)}^ \top })^{ - 1}}, \nonumber
\end{align}
which, in conjunction with \eqref{fg2}, leads to
\begin{align}
&{\Omega _1}\bigcap {{\Omega _2}}  \nonumber\\
&\subseteq \left\{\! {\sqrt 2 || {{R_{(k,p)}}\!{{( {{S} \!+\! {X_{(k,p)}}X_{(k,p)}^ \top })^{ - 0.5}}}} || \!>\! \beta \phi } \!\right\}\! \bigcap {{\Omega _2}}, \label{fg3}
\end{align}
where we denote
\begin{align}
S \triangleq ( {1 - \rho_{3} })M_{(k,p)}^{ - 2}. \label{kks}
\end{align}

We now define two additional events:
\begin{align}
&{\Phi _1} \!\triangleq\! \left\{\! {{{|| {{{( {S + {X_{(k,p)}}X_{(k,p)}^\top })^{ - 0.5}}}R_{(k,p)}^\top} ||^2}}} \right.\label{adf1}\\
&\hspace{1.2cm}\left. { > 16\mathfrak{c}{\kappa ^2}\ln ({{{({\det( {( {S \!+\! {X_{(k,p)}}X_{(k,p)}^ \top }){S^{ - 1}}})})^{0.5}}}\frac{1}{{{\delta _0}}}})} \!\right\}\!, \nonumber\\
&{\Phi _2}(\mathbf{u}) \!\triangleq\! \left\{\! {{{|| {{{( {S + {X_{(k,p)}}X_{(k,p)}^\top })^{ - 0.5}}}R_{(k,p)}^\top \mathbf{u}} ||}_{2}^2}} \right.\label{adf2}\\
&\hspace{1.3cm}\left. { > 4\mathfrak{c}{\kappa ^2}\ln (\!\! {{{({\det( {( {S \!+\! {X_{(k,p)}}X_{(k,p)}^\top }){S^{ - 1}}})})^{0.5}}}\frac{1}{{{\delta _0}}}})} \!\right\}\!, \nonumber
\end{align}
where $\mathbf{u} \in \mathcal{S}^{n-1}$.

We define $\breve{H}_{(k,p)} \triangleq  [ {\breve{H}_{k}, ~\breve{H}_{k+1}, ~\ldots, ~\breve{H}_{p-1}}]$, where
\begin{align}
\breve{H}_{k} \triangleq  \begin{cases}
[ {\mathbf{h}^{\mathbb{T}_k(1)}_{k},~\mathbf{h}^{\mathbb{T}_k(2)}_{k}, ~\ldots, ~\mathbf{h}^{\mathbb{T}_k(\left| {{\mathbb{T}_k}} \right|)}_{k}}], &\text{if}~\mathbb{T}_k \neq \varnothing\\
`\mathrm{null}',&\text{otherwise}.
\end{cases} \nonumber
\end{align}
Thus, $R_{(k,p)}$ in \eqref{defmatbdis2} equates to $R_{(k,p)} = \breve{H}_{(k+1,p)}X^{\top}_{(k,p)}$. As a consequence, $R_{(k,p)}^\top \mathbf{u} = X_{(k,p)}(\breve{H}^{\top}_{(k+1,p)}\mathbf{u})$.  We note that under Assumption \ref{asprv}-3), $\mathbf{u} \in \mathcal{S}^{n-1}$ implies $(\mathbf{h}^{m}_{k})^{\top}\mathbf{u}$ is $\mathcal{F}_{k}$-measurable and conditionally $\gamma$-sub-Gaussian for some $\gamma > 0$. We note that \eqref{kp1co} implies $4\mathfrak{c}{\kappa^2} = 2{\gamma^2}$. In light of  Lemma \ref{ckpq},  we have $\mathbf{P}[{\Phi _2}(\mathbf{u})] \leq \delta_{0}$. Furthermore, applying \eqref{aalem1}, with the setting of $\varepsilon = \frac{1}{2}$ in Lemma \ref{cko}, we obtain
\begin{align}
\mathbf{P}[{\Phi _1}] \leq {5^n}\mathop {\max }\limits_{\mathbf{u} \in \mathcal{N}} \mathbf{P}\left[ {{\Phi_2(\mathbf{u})}} \right] \le {5^n}{\delta _0}.\label{fg4}
\end{align}

We let  ${\delta _0} = \frac{\delta }{{2 \cdot {5^n}}}$, such that
\begin{align}
\beta  &\ge \frac{{4\sqrt {2\mathfrak{c}} \kappa }}{\phi }\sqrt {\ln \left(\frac{{2 \cdot {5^{1.5n}}}}{\delta}{{\left( {\frac{{1 - \rho_{3} }}{{10}}} \right)}^{0.5n}}\right)} \nonumber\\
 &\!=\! \frac{{4\sqrt {2\mathfrak{c}} \kappa }}{\phi }\sqrt {\ln \left(\frac{{2 \cdot {5^{1.5n}}}}{{2 \cdot {5^n} \cdot {\delta _0} \cdot {5^{0.5n}}}}{{\left( {\frac{{1 - \rho_{3} }}{2}} \right)}^{0.5n}}\right)} \nonumber\\
 &\!=\! \frac{{4\sqrt {2\mathfrak{c}} \kappa }}{\phi }\sqrt {\ln \left( {\frac{1}{{{\delta _0}}}{{\left( {\det \left( {\frac{{2{\mathbf{I}_n}}}{{1 - \rho_{3} }}} \right)} \right)}^{ - 0.5}}} \right)} \nonumber\\
 &\!=\! \frac{{4\sqrt {2\mathfrak{c}} \kappa }}{\phi }\sqrt {\ln \left( {\frac{1}{{{\delta _0}}}{{\left( {\det \left( {\left( {S + \frac{{1 + \rho_{3} }}{{1 - \rho_{3} }}S} \right){S^{ - 1}}} \right)} \right)}^{ - 0.5}}} \right)} \nonumber\\
 &\!\ge\! \frac{{4\sqrt {2\mathfrak{c}} \kappa }}{\phi }\!\sqrt {\ln \!\left(\! {\frac{1}{{{\delta _0}}}\!{{\left( {\det({\!( {S \!+\! {X_{(k,p)}}X_{(k,p)}^ \top }){S^{ - 1}}}\!)} \!\right)^{ - 0.5}}}} \!\right)}\!,\label{pq}
\end{align}
where the last inequality from its previous step is obtained via considering the inequality ${X_{(k,p)}}X_{(k,p)}^ \top  \le \frac{{1 + \rho_{3} }}{{1 - \rho_{3} }}S$ that follows from \eqref{kks} and \eqref{fg1}.

Combining the inequality in \eqref{fg3} with \eqref{pq} yields
\begin{align}
&|| {{R_{(k,p)}}{{( {S + {X_{(k,p)}}X_{(k,p)}^ \top })^{ - 0.5}}}} || \nonumber\\
& \!>\! \frac{{\beta \phi }}{{\sqrt 2 }} \!\ge\! 4\sqrt \mathfrak{c} \kappa \sqrt {\ln ( {\frac{1}{{{\delta _0}}}{{( {\det(( {S \!+\! {X_{(k,p)}}X_{(k,p)}^ \top ){S^{ - 1}}})})^{ - 0.5}}}})}\!, \nonumber
\end{align}
by which, and considering \eqref{fg3} and \eqref{adf1}, we deduce that under the condition \eqref{pq} if the event $\Omega _1$ occurs, the event ${\Phi _1}$ occurs consequently. We thus obtain:
\begin{align}
\mathbf{P}[ {{\Omega _1}\bigcap {{\Omega _2}} }]  \leq \mathbf{P}[ {{{\Phi _1}}\bigcap {{\Omega _2}} }].\label{fg4pk}
\end{align}

We note that the condition \eqref{fg4a} is equivalent to
\begin{align}
\lambda _{\min }^{0.5}( {( {1 \!-\! \rho_{3} }){\Gamma _{(k,p)}}}) \!\ge\! \frac{{4\sqrt {2\mathfrak{c}} \kappa }}{\phi }\!\sqrt {\ln ( {\frac{{2 \cdot {5^{1.5n}}}}{\delta }{{( {\frac{{1 - \rho_{3} }}{{10}}})^{0.5n}}}})}, \nonumber
\end{align}
inserting the definition of $\beta$ in \eqref{fg1j0} into which, we arrive at
\begin{align}
\beta  \ge \frac{{4\sqrt {2\mathfrak{c}} \kappa }}{\phi }\!\sqrt {\!\ln ({\frac{{2 \cdot {5^{1.5n}}}}{\delta }{{({\frac{{1 - \rho_{3} }}{{10}}})^{0.5n}}}})}, \nonumber
\end{align}
by which we conclude that \eqref{pq} holds if the condition \eqref{fg4a} is satisfied. Moreover, recalling that the event ${\Omega _2}$ always occurs under the condition \eqref{fg0aa} (proved in \underline{Upper Bound on $\mathbf{P}( {\Omega _2^\mathrm{c}})$}), we conclude from \eqref{fg4pk} and \eqref{fg4} that
\begin{align}
\mathbf{P}[ {{\Omega _1}\bigcap {{\Omega _2}} }]  \leq \mathbf{P}[ {\Phi _1}] \leq {5^n}{\delta _0} \nonumber
\end{align}
holds as long as both \eqref{fg0aa} and \eqref{fg4a} are satisfied. In addition, due to ${\delta _0} = \frac{\delta }{{2 \cdot {5^n}}}$, we have
\begin{align}
\mathbf{P}[ {{\Omega _1}\bigcap {{\Omega _2}} }]  \leq  \frac{\delta}{2}. \label{fg4pk2}
\end{align}
Combining \eqref{kp1} with \eqref{kpp1} and \eqref{fg4pk2} straightforwardly yields $\mathbf{P}[||A_{\mathrm{if}} - A|| > \phi] \leq \delta$.

\subsection{Proof of Statement 2)} We first note that
\vspace{-0.2cm}
\begin{align}
{( {||\mathbf{q}|| - ||\widehat{\mu}||})^2} - \underline{\mathfrak{h}} &\le ||\mathbf{q} - \widehat{\mu} |{|^2} - \underline{\mathfrak{h}} \le \left| {||\mathbf{q} - \widehat{\mu} |{|^2} - \underline{\mathfrak{h}}} \right|,\nonumber\\
{( {||\mathbf{r}|| - \sqrt{n}\mu})^2} - \overline{\mathfrak{h}} &\le ||\mathbf{r} - \mu {\mathbf{1}_n}|{|^2} - \overline{\mathfrak{h}} \le \left| {||\mathbf{r} - \mu {\mathbf{1}_n}|{|^2} - \overline{\mathfrak{h}}} \right|,\nonumber
\end{align}
which, respectively, imply that
\begin{align}
\mathbf{P}\!\left[{||\mathbf{q}|| - ||\widehat{\mu}||})^2 - \underline{\mathfrak{h}} \leq  {\rho _2}\right] &\geq \mathbf{P}\!\left[\left| {||\mathbf{q} - \widehat{\mu} |{|^2} - \underline{\mathfrak{h}}} \right| \leq  {\rho _2}\right],\nonumber\\
\mathbf{P}\!\left[{||\mathbf{r}|| - \sqrt{n}\mu})^2 - \overline{\mathfrak{h}} \leq  {\rho _1}\right] &\geq \mathbf{P}\!\left[\left| {||\mathbf{r} - \mu {\mathbf{1}_n}|{|^2} - \overline{\mathfrak{h}}} \right| \leq  {\rho _1}\right].\nonumber
\end{align}
Meanwhile, we note that $({||\mathbf{q}|| - ||\widehat{\mu}||})^2 - \underline{\mathfrak{h}} \leq  {\rho _2}$ and $({||\mathbf{r}|| - \sqrt{n}\mu})^2 - \overline{\mathfrak{h}} \leq  {\rho _1}$ are, respectively, equivalent to
\begin{align}
-\sqrt{{\rho _2} + \underline{\mathfrak{h}}} + ||\widehat{\mu}|| &\leq ||\mathbf{q}||\leq \sqrt{{\rho _2} + \underline{\mathfrak{h}}} + ||\widehat{\mu}||,\nonumber\\
-\sqrt{{\rho _1} + \overline{\mathfrak{h}}} + \sqrt{n}\mu &\leq ||\mathbf{r}|| \leq \sqrt{{\rho _1} + \overline{\mathfrak{h}}} + \sqrt{n}\mu,\nonumber
\end{align}
which indicates that
\begin{align}
&\mathbf{P}[||\mathbf{q}||\leq \sqrt{{\rho _2} + \underline{\mathfrak{h}}} + ||\widehat{\mu}||] \geq \mathbf{P}[{||\mathbf{q}|| - ||\widehat{\mu}||})^2 - \underline{\mathfrak{h}} \leq  {\rho _2}],\nonumber\\
&\mathbf{P}[||\mathbf{r}|| \leq \sqrt{{\rho _1} + \overline{\mathfrak{h}}} + \sqrt{n}\mu] \geq \mathbf{P}[{||\mathbf{r}|| - \sqrt{n}\mu})^2 - \overline{\mathfrak{h}} \leq  {\rho _1}],\nonumber
\end{align}
by which, we thus conclude that
\begin{align}
&\mathbf{P}[|\mathbf{q}||\leq \sqrt{{\rho _2} + \underline{\mathfrak{h}}} + ||\widehat{\mu}||] \geq \mathbf{P}[\left| {||\mathbf{q} - \widehat{\mu} |{|^2} - \underline{\mathfrak{h}}} \right| \leq  {\rho _2}],\label{mmk1}\\
&\mathbf{P}[|\mathbf{r}|| \!\leq\! \sqrt{{\rho _1} \!+\! \overline{\mathfrak{h}}} \!+\! \sqrt{n}\mu] \!\geq\! \mathbf{P}[\left| {||\mathbf{r} \!-\! \mu {\mathbf{1}_n}|{|^2} \!-\! \overline{\mathfrak{h}}} \right| \!\leq\!  {\rho _1}].\label{mmk2}
\end{align}

We note that $\widehat{\phi}  = ( {\sqrt {{\rho_1} + \overline{\mathfrak{h}}}  + \sqrt n \mu })\phi  + {\sqrt {{\rho_2} + \underline{\mathfrak{h}}}  + \left\| \widehat{\mu}  \right\|}$ in conjunction with \eqref{ckkoba1}, \eqref{mmk1} and \eqref{mmk2} results in
\begin{align}
&\mathbf{P}[|| {{\alpha_{\text{if}}} - \alpha }|| \leq \widehat{\phi} ] \nonumber\\
&\geq \mathbf{P}[\left\| {A - A_{\text{if}}} \right\|\left\| \overline{\mathbf{r}} \right\| + \left\| \mathbf{q} \right\| \leq  \widehat{\phi}]\nonumber\\
&\geq \mathbf{P}[\left\| {A - A_{\text{if}}} \right\| \leq \phi]\cdot\mathbf{P}[\left| {||\mathbf{q} - \widehat{\mu} |{|^2} - \underline{\mathfrak{h}}} \right| \leq  {\rho _1}] \nonumber\\
&\hspace{3.7cm}\cdot\mathbf{P}[\left| {||\mathbf{r} - \mu {\mathbf{1}_n}|{|^2} - \overline{\mathfrak{h}}} \right| \leq  {\rho _2}], \nonumber\\
& \geq 4\cdot(1 - \delta)\cdot{e^{{ - \frac{1}{{\mathfrak{c}{\kappa ^2}}}\min \left\{ {\frac{{\rho _2^2}}{{n\overline{\mathcal{C}}}},~{\rho _2}} \right\}}}}\cdot{e^{-{\frac{1}{{\mathfrak{c}{\kappa ^2}}}\!\min \left\{ {\frac{{\rho _1^2}}{{n\underline{\mathcal{C}}}},~{\rho _1}}\right\}}}},\nonumber
\end{align}
where the last inequality is obtained via considering $\mathbf{P} (||A-A_{\mathrm{if}}|| \leq \phi) \geq 1 - \delta$, \eqref{hhhccc2} and \eqref{hhhccc3}.

\section*{Appendix H: Proof of Proposition \ref{pop2}}
For $m > 2k- 1$, the third term in the right-hand side of \eqref{dv1} can be rewritten as
\begin{align}
\sum\limits_{i = k - 1}^{m - 2}\!\!\!\!{{A^i}({\bf{f}}(m \!-\! 1 \!-\! i)}\!+\!\mathbf{a})  \!=\! &\sum\limits_{i = m - k}^{m - 2}\!\!\!\!\!{{A^i}({\bf{f}}(m \!-\! 1 \!-\! i)\!+\!\mathbf{a})} \nonumber\\
&+ \sum\limits_{i = k - 1}^{m - k-1}\!\!\!\!\!{{A^i}({\bf{f}}(m \!-\! 1 \!-\! i)\!+\!\mathbf{a})}. \label{rre}
\end{align}

Under Assumption \ref{asprv}-1), it is straightforward to verify from \eqref{dv1} and \eqref{rre} that
\begin{align}
&{\bf{E}}[\mathbf{r}_k^m(\mathbf{r}_k^m)^{\top}] \nonumber\\
&\geq ( {{A^{k - 1}} - {A^{m - 1}}})\mathbf{E}[ {{\bf{x}}(1){{\bf{x}}^\top}(1)}]{( {{A^{k - 1}} - {A^{m - 1}}})^\top} \nonumber\\
&~~~~+ \mathbf{E}[{( {\sum\limits_{i = k - 1}^{m - k-1}\!\!\! {{A^i}{\bf{f}}(m \!-\! 1 \!-\! i)} }){{( {\sum\limits_{i = k - 1}^{m - k-1} \!\!\!{{A^i}{\bf{f}}(m \!-\! 1 \!-\! i)} } )^\top}}}] \nonumber\\
&~~~~+ \mathbf{E}[ {{\bf{w}}_k^m{{( {{\bf{w}}_k^m})^\top}}}] + \mathbf{E}[ {( {\sum\limits_{i = k - 1}^{m - 2} {{A^i}{\bf{a}}} }){{( {\sum\limits_{i = k - 1}^{m - 2} {{A^i}{\bf{a}}} })^\top}}}]\nonumber\\
&\geq \left( {{A^{k - 1}} - {A^{m - 1}}} \right)\!\!{\left( {{A^{k - 1}} - {A^{m - 1}}} \right)^\top}\sigma_{\mathrm{i}}^2 + 2\sigma_{\mathrm{o}}^2{\mathbf{I}_n} \nonumber\\
&~~~~+ \sum\limits_{i = k - 1}^{m - k-1} \!\!\!{{A^i}{{( {{A^i}})^\top}}\sigma_{\mathrm{p}}^2}.\label{or}
\end{align}

Considering $m > k$, we have
\begin{align}
&\left( {{A^{k - 1}} - {A^{m - 1}}} \right){\left( {{A^{k - 1}} - {A^{m - 1}}} \right)^ \top } \nonumber\\
& = {A^{k - 1}}\left( {{\mathbf{I}_n} - {A^{m - k}}} \right){\left( {{\mathbf{I}_n} - {A^{m - k}}} \right)^ \top }{\left( {{A^{k - 1}}} \right)^\top}. \label{cmgh}
\end{align}

For a square matrix $A$, we known that ${\sigma _i}\left( A \right) = \sqrt {{\lambda _i}\left( {A{A^\top}} \right)}$, where $\lambda_{i}\left(A\right)$ denotes the $i$th eigenvalue of $A$. We thus have
\begin{align}
\left( {{\mathbf{I}_n} \!-\! {A^{m - k}}} \right){\left( {{\mathbf{I}_n} \!-\! {A^{m - k}}} \right)^ \top } \!\ge\! \mathop {\min }\limits_{i \in \left\{ {1, \ldots ,n} \right\}} \!\!\left\{\! {{{\sigma^2 _i( {A^{m - k} - \mathbf{I}_{n}} )}}} \!\right\}{\mathbf{I}_n}, \nonumber
\end{align}
which, in conjunction with \eqref{cmgh}, \eqref{hx1} and \eqref{hx2}, leads to
\begin{align}
\left( {{A^{k - 1}} \!-\! {A^{m - 1}}} \right)\!{\left( {{A^{k - 1}} \!-\! {A^{m - 1}}} \right)^ \top } \ge \widehat{\underline{\sigma}}_{A}^{2k - 2}\widetilde{\underline{\sigma}}_{A}^{2}\mathbf{I}_{n}. \label{cmgh1}
\end{align}
We note that
\begin{align}
\sum\limits_{i = k - 1}^{m - k-1} \!\!\! {A^i}{\left( {{A^i}} \right)^ \top }\sigma _{\rm{p}}^2 \ge \sum\limits_{i = k - 1}^{m - k-1} \!\!\! \widehat{\underline{\sigma}}_{A}^{2i}\sigma _{\rm{p}}^2{\mathbf{I}_n} \geq \widehat{\underline{\sigma}}_{A}^{2k-2}\sigma _{\rm{p}}^2{\mathbf{I}_n}, \nonumber
\end{align}
inserting which along with \eqref{cmgh1} into \eqref{or} yields ${\bf{E}}\!\left[\!\mathbf{r}_k^m(\mathbf{r}_k^m)^{\top}\!\right] \geq \mathfrak{f}_{1}(A)\mathbf{I}_{n}$, where $\mathfrak{f}_{1}(A)$ is given in \eqref{defho}.

For $m \leq 2k- 1$, we can straightforwardly obtain from \eqref{dv1}:
\begin{align}
{\bf{E}}\!\left[\!\mathbf{r}_k^m(\mathbf{r}_k^m)^{\top}\!\right]
&\!\geq\! \left(\! {{A^{k - 1}} \!-\! {A^{m - 1}}} \!\right)\!\mathbf{E}\!\left[ {{\bf{x}}(1){{\bf{x}}^\top}\!(1)} \right]\!\!{\left(\! {{A^{k - 1}} \!-\! {A^{m - 1}}} \!\right)\!\!^\top} \nonumber\\
&~~+ \mathbf{E}\!\left[ {{\bf{w}}_k^m{{\left( {{\bf{w}}_k^m} \right)^\top}}} \right],\nonumber\\
&\!=\! \left( {{A^{k - 1}} \!-\! {A^{m - 1}}} \right)\!\!{\left( {{A^{k - 1}} \!-\! {A^{m - 1}}} \right)^\top}\sigma_{\mathrm{i}}^2 \!+\! 2\sigma_{\mathrm{o}}^2{\mathbf{I}_n}. \nonumber
\end{align}
Following the same steps of obtaining \eqref{cmgh1}, we have ${\bf{E}}\left[\mathbf{r}_k^m(\mathbf{r}_k^m)^{\top}\right] \geq (\widehat{\underline{\sigma}}_{A}^{2k - 2}\widetilde{\underline{\sigma}}_{A}^{2}\sigma_{\mathrm{i}}^2 + 2\sigma_{\mathrm{o}}^2)\mathbf{I}_{n} = \mathfrak{f}_{2}(A)\mathbf{I}_{n}$, where $\mathfrak{f}_{2}(A)$ is given in \eqref{defho2}.

\section*{Appendix I: Proof of Proposition \ref{pppt}}
Considering the matrix ${{\Upsilon}_{({k},{p})}}$ given in \eqref{gh9aa}, we have
\begin{align}
 {|| {{\Upsilon _{(k,p)}}} ||}  = {|| {{M_{(k,p)}}} ||}. \label{cco1a}
\end{align}
Considering $||\mathbf{I}_{\sum\limits_{m = k}^{p - 1} \!\!{\left| {{\mathbb{T}_m}} \right|n}}|| = 1$, we obtain from \eqref{ghmatrix} that
\begin{align}
|| {{\Pi _{(k,p)}}} || \leq || \widehat{\mathcal{A}}_{(k,p)} || + || \widetilde{\mathcal{A}}_{( {k,p})} || + || \breve{\mathcal{A}}_{(k,p)} || + 1. \label{cco1b}
\end{align}

With the knowledge of upper bound $\overline{\widetilde{\sigma}}_{A}$ in \eqref{hx3}, it is straightforward to obtain from \eqref{gh4}-\eqref{gh5} that
\begin{align}
|| \breve{\mathcal{A}}_{( {k,p})} || \leq \mathop {\max }\limits_{q \in \left\{ {k, \ldots ,p - 1} \right\}}\left\{ {\frac{{{\overline{\widetilde{\sigma}}_A} - \overline{\widetilde{\sigma}}_A^{q - 1}}}{{1 - {\overline{\widetilde{\sigma}}_A}}}}\right\}. \label{cv1}
\end{align}
Meanwhile, we obtain from \eqref{gh5bba}-\eqref{gh7} that
\begin{align}
|| \widetilde{\mathcal{A}}_{( {k,p})} || \leq \mathop {\max }\limits_{j < m \in \left\{ {k, \ldots ,p - 1} \right\}}\left\{{\frac{{\overline{\widetilde{\sigma}}_A^{j - 1} -\overline{\widetilde{\sigma}}_A^{m - 1}}}{{1 - {\overline{\widetilde{\sigma}}_A}}}}\right\}.\label{cv2}
\end{align}

Noticing the upper bound $\overline{\widehat{\sigma}}$ in \eqref{hx4}, we obtain from \eqref{gh1}-\eqref{gh3} that $|| \widehat{\mathcal{A}}_{(k,p)}|| \leq \overline{\widehat{\sigma}}$, substituting which together with \eqref{cv1} and \eqref{cv2} into \eqref{cco1b}, we arrive at $\left\| {{\Pi _{(k,p)}}} \right\| \leq \mathfrak{g}_{(k,p)}$, where $\mathfrak{g}_{(k,p)}$ is defined in \eqref{cvop}. Furthermore, with well-known inequality $\left\| {GH} \right\| \le \left\| G \right\|\left\| H \right\|$, it follows from \eqref{cco1a}  that
\begin{align}
||{{\Pi}^{\top}_{({{k},{p}})}}\!{{\Upsilon}_{({k},{p})}}|| \!\leq\! ||{{\Pi}^{\top}_{({{k},{p}})}}||||{{\Upsilon}_{({k},{p})}}|| \!\leq\!{|| {{M_{(k,p)}}} ||}{\mathfrak{g}_{(k,p)}}.\label{ckmq}
\end{align}
Noting $\mathfrak{f}_{(k,m)}$ \eqref{ckjdef}, and substituting \eqref{ckj} into \eqref{kop3p} yields
\begin{align}
M_{(k,p)}^{ - 2} \ge \sum\limits_{r = k}^{p - 1} {\sum\limits_{q = {\mathbb{T}_r}(1)}^{{\mathbb{T}_r}( {\left| {{\mathbb{T}_r}} \right|})} {{\mathfrak{f}_{({r,q})}}\mathbf{I}_{n}} }. \nonumber
\end{align}
As a consequence, we have
\begin{align}
\left\|M_{(k,p)}\right\|^{2} \leq \frac{{{1}}}{{\sum\limits_{r = k}^{p - 1} {\sum\limits_{q = {\mathbb{T}_r}(1)}^{{\mathbb{T}_r}( {\left| {{\mathbb{T}_r}} \right|})} {{\mathfrak{f}_{( {r,q})}}} } }}, \nonumber
\end{align}
inserting which into \eqref{ckmq}  leads to \eqref{aah}.

\bibliographystyle{IEEEtran}
\bibliography{ref}

\end{document}